\documentclass[english]{article}
\usepackage[T1]{fontenc}
\usepackage[latin9]{inputenc}
\usepackage{geometry}
\usepackage{xcolor}
\usepackage{babel}
\usepackage{amsmath}
\usepackage{graphicx}
\usepackage[authoryear]{natbib}
\PassOptionsToPackage{normalem}{ulem}
\usepackage{ulem}
\usepackage{subscript}
\usepackage[unicode=true,
 bookmarks=true,bookmarksnumbered=false,bookmarksopen=false,
 breaklinks=true,pdfborder={0 0 0},backref=false,colorlinks=true]
 {hyperref}
\hypersetup{pdftitle={Dynamical transitions in a pollination--herbivory interaction},
 pdfauthor={Tomas Revilla and Francisco Encinas-Viso},
 pdfkeywords={mutualism, pollination, herbivory, insects, stage-structure, oscillations}}
\usepackage{breakurl}

\makeatletter

\providecommand{\tabularnewline}{\\}
\providecolor{lyxadded}{rgb}{0,0,1}
\providecolor{lyxdeleted}{rgb}{1,0,0}
\newenvironment{lyxcode}
{\par\begin{list}{}{
\setlength{\rightmargin}{\leftmargin}
\setlength{\listparindent}{0pt}% needed for AMS classes
\raggedright
\setlength{\itemsep}{0pt}
\setlength{\parsep}{0pt}
\normalfont\ttfamily}%
 \item[]}
{\end{list}}

\usepackage[auth-lg]{authblk}
   \author[1]{Tomás A. Revilla}
   \author[2]{Francisco Encinas-Viso}
   \affil[1]{Centre for Biodiversity Theory and Modelling, 09200 Moulis, France. 
                    Email:tomrevilla@gmail.com}
   \affil[2]{CSIRO Plant Industry, GPO Box 1600, Canberra, ACT 2601, Australia. 
                    Email:francisco.encinas.viso@csiro.au}

\usepackage{lineno}
\modulolinenumbers[2]

\makeatother

\begin{document}

\title{Dynamical transitions in a pollination--herbivory interaction: a
conflict between mutualism and antagonism}

\maketitle

\begin{abstract}
Plant-pollinator associations are often seen as purely mutualistic,
while in reality they can be more complex. Indeed they may also display
a diverse array of antagonistic interactions, such as competition
and victim--exploiter interactions. In some cases mutualistic and
antagonistic interactions are carried-out by the same species but
at different life-stages. As a consequence, population structure affects
the balance of inter-specific associations, a topic that is receiving
increased attention. In this paper, we developed a model that captures
the basic features of the interaction between a flowering plant and
an insect with a larval stage that feeds on the plant's vegetative
tissues (e.g. leaves) and an adult pollinator stage. Our model is
able to display a rich set of dynamics, the most remarkable of which
involves victim--exploiter oscillations that allow plants to attain
abundances above their carrying capacities, and the periodic alternation
between states dominated by mutualism or antagonism. Our study indicates
that changes in the insect's life cycle can modify the balance between
mutualism and antagonism, causing important qualitative changes in
the interaction dynamics. These changes in the life cycle could be
caused by a variety of external drivers, such as temperature, plant
nutrients, pesticides and changes in the diet of adult pollinators.

Keywords: \emph{mutualism, pollination, herbivory, insects, stage-structure,
oscillations}
\end{abstract}

\section{Introduction}
\begin{quote}
\begin{flushleft}
\emph{Il faut bien que je supporte deux ou trois chenilles si je veux
connaître les papillons}
\par\end{flushleft}

\begin{flushleft}
Le Petit Prince, Chapitre IX -- Antoine de Saint-Exupéry
\par\end{flushleft}
\end{quote}
Mutualism entails costs in addition to benefits. Conflicting goals
can lead to cheating where one party incurs the cost of providing
energy to enable mutualism, while the other exploits, but does not
reciprocate (e.g. nectar robbers). There can also be costs concerning
other detrimental interactions that run in parallel with mutualism,
such as predation, parasitism or competition involving the same parties.
Moreover, some of these antagonistic interactions (e.g. competition)
seem to be important for the evolution and stability of mutualism
\citep{Jones_compmut_2012}. In general, these costs have important
consequences at the population and community level, because the net
outcome can turn out beneficial or detrimental, but perhaps more interestingly,
variable \citep{bronstein-tree94}. Variable interactions challenge
the view that ecological communities are structured by well defined
interactions at the species level such as competition (--,--), victim-exploiter
(--,+) or mutualism (+,+).

Pollination is one of the most important mutualisms occurring between
plants and animals. This form of trading resources for services greatly
explains the evolutionary success of flowering plants in almost all
terrestrial systems. It is responsible for the well being of ecosystem
services. During the larval stage of many insect pollinators, such
as Lepidopterans (butterflies and moths), the larvae feed on plant
leaves to mature and become adult pollinators \citep{adler_bronstein-ecology04,wackers_etal-are07,bronstein_etal-annbot09,altermatt_pearse-amnat11}.
These ontogenetic diet shifts \citep{rudolf_lafferty-ecolett11} are
very common and important in understanding the ecological and evolutionary
dynamics of plant--animal mutualisms. Interestingly, in some cases
larvae feed on the same plant species that they will pollinate as
adults \citep{irwin-cb10,bronstein_etal-annbot09}. This shows that
in several cases mutualistic and antagonistic interactions are exerted
by the same species. and a potential conflict arises for the plant.
between the benefits of mutualism and the costs of herbivory. One
of the best known examples is the interaction between tobacco plants
(\emph{Nicotiana} \emph{attenuata}) and the hawkmoth (\emph{Manduca}
\emph{sexta}) \citep{baldwin-oecologia88,kessler_etal-cb10}, whose
larva is commonly called the tobacco hornworm. There are other examples
of this type of interaction in the genus \emph{Manduca} (Sphingidae),
such as between the tomato plant (\emph{Lycopersicon esculentum})
and the five-spotted hawkmoth (\emph{Manduca quinquemaculata}) \citep{Kennedy_2003}.
These larvae have received a lot of attention due to their negative
effects on agricultural crops \citep{Campbell_1991_pests}.

The interaction between \emph{Manduca sexta} and \emph{Datura wrightii}
(Solanacea) \citep{bronstein_etal-annbot09,alarcon_etal-ecoent08}
is another good example illustrating the costs and benefits of pollination
mutualisms \citep{bronstein_etal-annbot09}. \emph{D. wrightii} provides
high volumes of nectar and seems to depend heavily on the pollination
service by \emph{M. sexta} adults \citep{alarcon_etal-ecoent08}.
However, \emph{M. sexta} larvae, which feed on \emph{D. wrightii}
vegetative tissue, can have severe negative effects on plant fitness
\citep{McFadden_1968,Barron-Gafford_2012}. We could assume that the
benefits of pollination might outweigh the costs of herbivory for
this mutualism to be relatively viable. The question is what are the
conditions, in terms of benefits (pollination) and costs (herbivory),
for this mutualistic interaction to be stable?

In the pollination--herbivory cases mentioned previously the benefits
and costs for the plant are clearly differentiated. This is because
the role of an insect as a pollinator or herbivore depends on the
stage in its life cycle \citep{miller_rudolf-tree11}. Thus, whether
mutualism or herbivory dominates the interaction is dependent on insect
abundance and its population structure. In other words the \emph{cost:benefit}
ratio must be positively related with the insect's \emph{larva:adult}
ratio. For a hypothetical scenario in which the costs of herbivory
(--) and the benefits of pollination (+) are balanced for the plant
(0), an increase in larval abundance relative to adults should bias
the relationship towards a victim-exploiter one (--,+). Whereas an
increase in adult abundance relative to larvae should bias the relationship
towards mutualism (+,+). Under equilibrium conditions, one would expect
transitions (bifurcations) from (--,+) to (0,+) to (+,+) and vice-versa
as relevant parameters affecting the plant and the insect life-histories
vary, such as flower production, mortalities or larvae maturation
rates. However, under dynamic scenarios the outcome may be more complex:
a victim--exploiter state (--,+) enhances larva development into pollinating
adults, but this tips the interaction into a mutualism (+,+), which
in turn contributes greater production of larva leading back to a
victim--exploiter state (--,+). This raises the possibility of feedback
between the plant--insect interaction and insect population structure,
which can potentially lead to periodic alternation between mutualism
and herbivory. Thus, when non-equilibrium dynamics are involved, questions
concerning the overall nature (positive, neutral or negative) of mixed
interactions may not have simple answers.

In this article we study the feedback between insect population structure,
pollination and herbivory. We want to understand how the balance between
costs (herbivory) and benefits (pollination) affects the interaction
between plants (e.g. \emph{D. wrightii}) and herbivore--pollinator
insects (e.g. \emph{M. sexta})? Also what role does insect development
have in this balance and on the resulting dynamics? We use a mathematical
model which considers two different resources provided by the same
plant species, nectar and vegetative tissues. Nectar consumption benefits
the plant in the form of fertilized ovules, and consumption of vegetative
tissues by larvae causes a cost. Our model predicts that the balance
between mutualism and antagonism, and the long term stability of the
plant--insect association, can be greatly affected by changes in larval
development rates, as well as by changes in the diet of adult pollinators.

\section{PLA (plant-larva-adult) model\label{sec:Model}}

Our model concerns the dynamics of the interaction between a plant
and an insect. The insect life cycle comprises an adult phase that
pollinates the flowers and a larval phase that feed on non-reproductive
tissues of the same plant. Adults oviposit on the same species that
they pollinate (e.g. \emph{D. wrightii -- M. sexta} interaction).
Let denote the biomass densities of the plant, the larva, and the
adult insect with $P,L$ and $A$ respectively. An additional variable,
the total biomass of flowers $F$, enables the mutualism by providing
resources to the insect (nectar), and by collecting services for the
plant (pollination). The relationship is \emph{facultative--obligatory}.
In the absence of the insect, the plant's vegetative biomass grows
logistically, preventing its extinction. In the absence of the plant
however, the insect always goes extinct because larval development
relies exclusively on herbivory, even if the adults pollinate other
plant species. This is based on the biology of \emph{M. sexta} \citep{bronstein_etal-annbot09}.
The mechanism of interaction between these four variables ($P,L,A,F$)
is described by the following system of ordinary differential equations
(ODE):

\begin{equation}
\begin{aligned}\frac{dP}{dt} & =rP(1-cP)+\sigma aFA-bPL\\
\frac{dF}{dt} & =sP-wF-aFA\\
\frac{dL}{dt} & =\epsilon aFA+gA-\gamma bPL-mL\\
\frac{dA}{dt} & =\gamma bPL-nA
\end{aligned}
\label{eq:pfla}
\end{equation}

\noindent where $r$: plant intrinsic growth rate, $c$: plant intra-specific
self-regulation coefficient (also the inverse its carrying capacity),
$a$: pollination rate, $b$: herbivory rate, $s:$ flower production
rate, $w$: flower decay rate, $m,n$: larva and adult mortality rates,
$\sigma$: plant pollination efficiency ratio, $\epsilon$: adult
consumption efficiency ratio. Like $\epsilon$, parameter $\gamma$
is also a consumption efficiency ratio, but we will call it the maturation
rate for brevity since we will refer to it frequently. Our model assumes
that pollination leads to flower closure \citep{primack-ares85},
causing resource limitation for adult insects. Parameter $g$ represents
a reproduction rate resulting from the pollination of other plants
species, which we do not model explicitly. Most of our results are
for $g=0$.

We now consider the fact that flowers are ephemeral compared with
the life cycles of plants and insects. In other words, some variables
$(P,L,A)$ have slower dynamics, and others $(F)$ are fast \citep{rinaldi_scheffer-ecosystems00}.
Given the near constancy of plants and animals in the flower equation
of (\ref{eq:pfla}), we can predict that flowers will approach a quasi-steady-state
(or quasi-equilibrium) biomass $F\approx sP/(w+aA)$, before $P,L$
and $A$ can vary appreciably. Substituting the quasi-steady-state
biomass in system (\ref{eq:pfla}) we arrive at:

\begin{equation}
\begin{aligned}\frac{dP}{dt} & =rP(1-cP)+\sigma\left[\frac{asA}{w+aA}\right]P-bPL\\
\frac{dL}{dt} & =\epsilon\left[\frac{asP}{w+aA}\right]A+gA-\gamma bPL-mL\\
\frac{dA}{dt} & =\gamma bPL-nA
\end{aligned}
\label{eq:upla}
\end{equation}

In system (\ref{eq:upla}) the quantities in square brackets can be
regarded as functional responses. Plant benefits saturate with adult
pollinator biomass, i.e. pollination exhibits diminishing returns.
The functional response for the insects is linear in the plant biomass,
but is affected by intraspecific competition \citep{schoener-tpb78}
for mutualistic resources.

We non-dimensionalized this model to reduce the parameter space from
12 to 9 parameters, by casting biomasses with respect to the plant's
carrying capacity $(1/c)$ and time in units of plant biomass renewal
time $(1/r)$. This results in a PLA (plant, larva, adult) scaled
model:

\begin{equation}
\begin{aligned}\frac{dx}{d\tau} & =x(1-x)+\sigma\frac{\alpha z}{\eta+z}x-\beta xy\\
\frac{dy}{d\tau} & =\epsilon\frac{\alpha x}{\eta+z}z+\phi z-\gamma\beta xy-\mu y\\
\frac{dz}{d\tau} & =\gamma\beta xy-\nu z
\end{aligned}
\label{eq:pla}
\end{equation}

Table \ref{tab:vars_and_pars} lists the relevant transformations.

\begin{table}
\protect\caption{\label{tab:vars_and_pars}Variables and parameters of the scaled PLA
model (\ref{eq:pla}) and values used for numerical analyses. The
last column shows a corresponding set of parameter values in the unscaled
version of the same model (\ref{eq:upla}), for plant carrying capacities
of $c^{-1}=100$ biomass units, and $r^{-1}=20$ time units.}

\centering{}%
\begin{tabular}{llcl}
\hline 
Symbol & Description & Value & $c=0.01,r=0.05$\tabularnewline
\hline 
\hline 
$x=cP,y=cL,z=cA$ & plant, larval and adult biomass & variable & \tabularnewline
$\tau=rt$ & time & variable & \tabularnewline
$\alpha=s/r$ & asymptotic pollination rate & 5 & $s=0.25$\tabularnewline
$\eta=wc/a$ & half-saturation constant of pollination & 0.1 & $w=0.5$ \& $a=0.05$\tabularnewline
$\beta=b/rc$ & herbivory rate & 0 to 100 & $b=0$ to 0.05\tabularnewline
$\mu=m/r$ & larva mortality rate & 1 & $m=0.05$\tabularnewline
$\nu=n/r$ & adult mortality rate & 2 & $n=0.1$\tabularnewline
$\phi=g/r$ & insect intrinsic reproduction rate & 0 or 1 & $g=0$ or 0.05\tabularnewline
$\sigma$ & plant pollination conversion ratio & 5 & \tabularnewline
$\epsilon$ & insect pollination conversion ratio & 0.5 & \tabularnewline
$\gamma$ & maturation rate (herbivory conversion ratio) & 0 to 0.1 & \tabularnewline
\hline 
\end{tabular}
\end{table}

There is an important clarification to make concerning the nature
and scales of the conversion efficiency ratios $\sigma,\epsilon$
involved in pollination, and $\gamma$ for herbivory and maturation.
This has to do with the fact that flowers \emph{per se} are not resources
or services, but \emph{organs} that enable the mutualism to take place,
and they mean different things in terms of biomass production for
plants and animals. For insects, the yield of pollination is thermodynamically
constrained. First of all, a given biomass $F$ of flowers contains
an amount of nectar that is necessarily less than $F$. More importantly,
part of this nectar is devoted to survival, or wasted, leaving even
less for reproduction. Similarly, not all the biomass consumed by
larvae will contribute to their maturation to adult. \emph{Ergo} $\epsilon<1,\gamma<1$.
Regarding the returns from pollination for the plants, the situation
is very different. Each flower harbors a large number of ovules, thus
a potentially large number of seeds \citep{fagan_etal-theorecol14},
each of which will increase in biomass by consuming resources not
considered by our model (e.g. nutrients, light). Consequently, a given
biomass of pollinated flowers can produce a larger biomass of mature
plants, making $\sigma$ larger than 1.

\section{Results\label{sec:Results}}

The PLA model (\ref{eq:pla}) has many parameters, however here we
focus on herbivory rates $(\beta)$ and larvae maturation $(\gamma)$,
because increasing $\beta$ turns the net balance interaction towards
antagonism, whereas increasing $\gamma$ shifts insect population
structure towards the adult phase, turning the net balance towards
mutualism. Both parameters also relate to the state variables at equilibrium
(i.e. $z/y=\beta\gamma x/\nu$ in (\ref{eq:pla} for $dz/d\tau=0$).
In section \ref{sub:numerical} we studied the joint effects of varying
$\beta$ and $\gamma$ numerically (parameter values in Table \ref{tab:vars_and_pars}).
In section \ref{sub:analytical} we present a simplified graphical
analysis of our model, in order to explain how different dynamics
can arise, by varying $\beta,\gamma$ and other parameters.

\subsection{Numerical results\label{sub:numerical}}

\begin{figure}
\begin{centering}
\includegraphics[angle=-90,width=0.6\paperwidth]{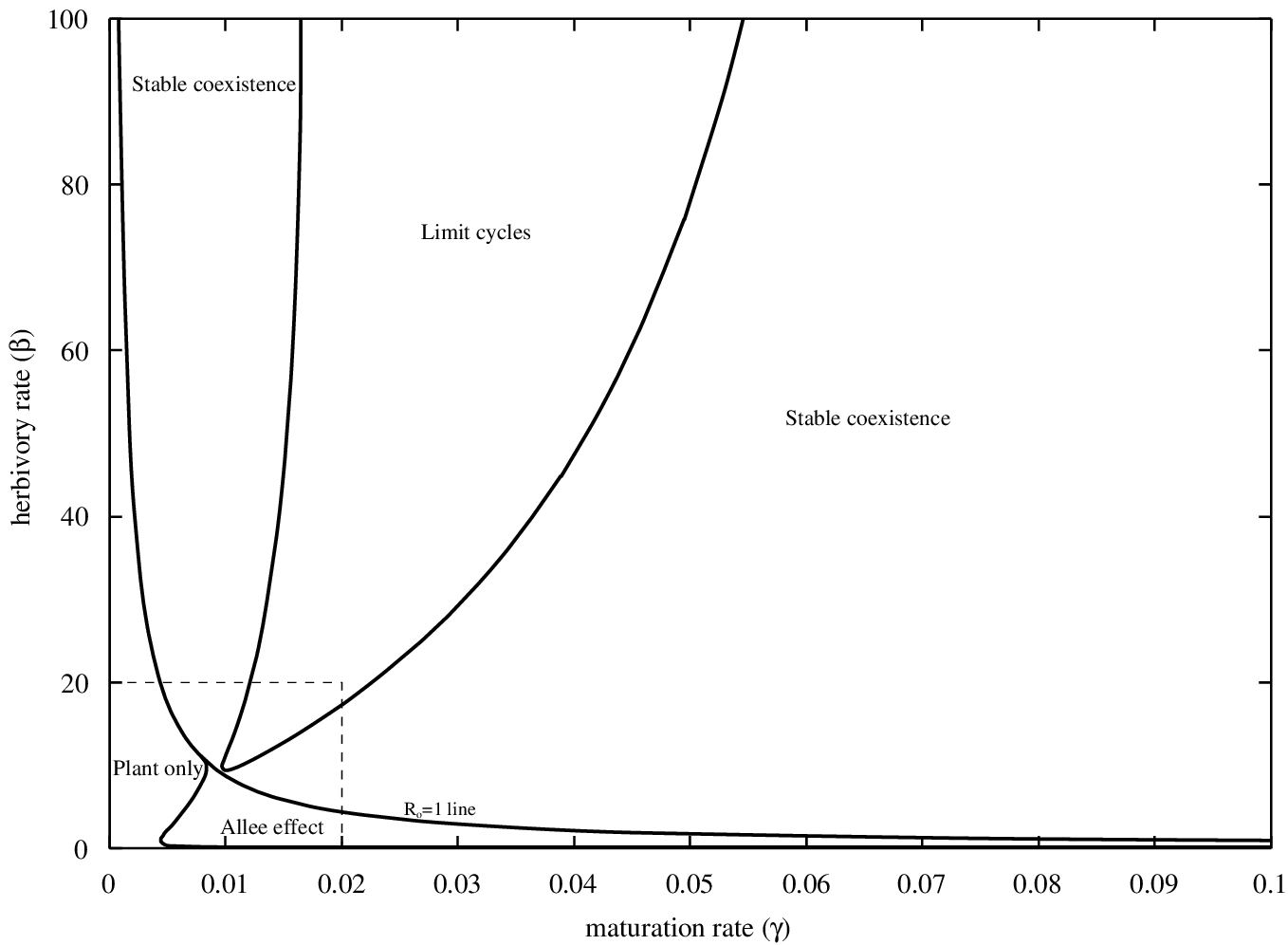}
\par\end{centering}

\protect\caption{\label{fig:beta_vs_gamma_phi0} Outcomes of the PLA model as a function
of the larval maturation and herbivory rates for specialist pollinators
$(\phi=0)$. The rectangular region in the bottom left is analyzed
with more detail in Appendix A.}
\end{figure}

Figure \ref{fig:beta_vs_gamma_phi0} shows interaction outcomes of
the PLA model, as a function of $\beta$ and $\gamma$ for specialist
pollinators $(\phi=0)$. This parameter space is divided by a decreasing
$R_{o}=1$ line that indicates whether or not insects can invade when
rare. $R_{o}$ is defined as (see derivation in Appendix A):

\begin{equation}
R_{o}=\frac{\epsilon\alpha\gamma\beta}{\eta\nu(\mu+\gamma\beta)}\label{eq:r0}
\end{equation}

\noindent and we call it the \emph{basic reproductive number}, according
to the argument that follows. Consider the following in system (\ref{eq:pla}):
if the plant is at carrying capacity $(x=1)$, and is invaded by a
very small number of adult insects $(z\approx0)$, the average number
of larvae produced by a single adult in a given instant is $\epsilon\alpha x/(\eta+z)\approx\epsilon\alpha/\eta$,
and during its life-time $(\nu^{-1})$ it is $\epsilon\alpha/\eta\nu$.
Larvae die at the rate $\mu$, or mature with a rate equal to $\gamma\beta x=\gamma\beta$,
per larva. Thus, the probability of larvae becoming adults rather
than dying is $\gamma\beta/(\mu+\gamma\beta)$. Multiplying the life-time
contribution of an adult by this probability gives the expected number
of new adults replacing one adult per generation during an invasion
($R_{o}$). More formally, $R_{o}$ is the expected number of adult-insect-grams
replacing one adult-insect-gram per generation (assuming a constant
mass-per-individual ratio).

Below the $R_{o}=1$ line, small insect populations cannot replace
themselves $(R_{o}<1)$ and two outcomes are possible. If the maturation
rate is too low, the plant only equilibrium $(x=1,y=z=0)$ is globally
stable and plant--insect coexistence is impossible for all initial
conditions. If the maturation rate is large enough, stable coexistence
is possible, but only if the initial plant and insect biomass are
large enough. This is expected in models where at least one species,
here the insect, is an obligate mutualist. In this region of the space
of parameters, the growth of small insect populations increases with
population size, a phenomenon called the Allee effect \citep{stephens_etal-oikos99}.

Above the $R_{o}=1$ line the plant only equilibrium is always unstable
against the invasion of small insect populations $(R_{o}>1)$. Plants
and insects can coexist in a stable equilibrium or via limit cycles
(stable oscillations). The zone of limit cycles occurs for intermediate
values of the maturation rate ($\gamma$) and it widens with rate
of herbivory ($\beta$).

Plant equilibrium when coexisting with insects can be above or below
the carrying capacity $(x=1)$. When above carrying capacity the net
result of the interaction is a mutualism (+,+). While in the second
case we have antagonism, more specifically net herbivory (--,+). As
it would be expected, increasing herbivory rates $(\beta)$ shifts
this net balance towards antagonism (low plant biomass), while decreasing
it shifts the balance towards mutualism (high plant biomass). The
quantitative response to increases in the maturation rate $(\gamma)$
is more complex however (see the bifurcation plot in Appendix A).

\begin{figure}
\begin{centering}
\includegraphics[clip,width=0.6\paperwidth]{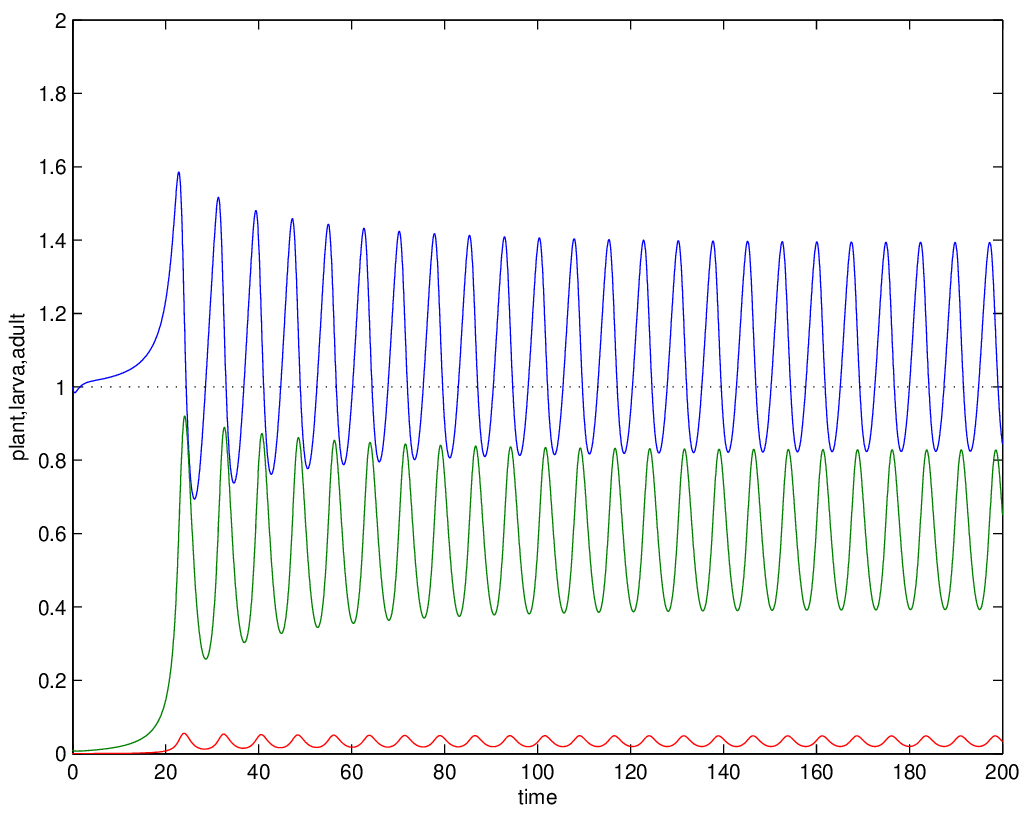}
\par\end{centering}

\protect\caption{\label{fig:dynamics_muther}Limit cycles in the PLA model (\ref{eq:pla})
with plant biomasses alternating above and below the carrying capacity
(dotted line). Parameters as in Table \ref{tab:vars_and_pars}, with
$\gamma=0.01,\beta=10$. Blue:plant, green:larva, red:adult.}
\end{figure}

Given that there is herbivory, we encounter victim--exploiter oscillations.
However, the oscillations in the PLA model are special in the sense
that the plant can attain maximum biomasses above the carrying capacity
$(x>1)$. For an example see Figure \ref{fig:dynamics_muther}. Instead
of a stable balance between antagonism and mutualism, we can say that
the outcome in Figure \ref{fig:dynamics_muther} is a periodic alternation
of both cases. This is not seen in simple victim--exploiter models,
where oscillations are always below the victim's carrying capacity
\citep{rosenzweig_macarthur-amnat63,rosenzweig-science71}. The relative
position of the cycles along the plant axis is also affected by herbivory:
if $\beta$ decreases (increases), plant maxima and minima will increase
(decrease) in Figure \ref{fig:dynamics_muther} (see bifurcation plot
in Appendix A). In some cases the entire plant cycle (maxima and minima)
ends above the carrying capacity if $\beta$ is low enough (see Appendix
C), but further decrease causes damped oscillations. We also found
examples in which coexistence can be stable or lead to limit cycles
depending on the initial conditions (see example in Appendix C), but
this happens in a very restrictive region in the space of parameters
(see bifurcation plot in Appendix A). Limit cycles can also cross
the plant's carrying capacity under the original interaction mechanism
(\ref{eq:pfla}), which does not assume the steady--state in the flowers
(see an example in Appendix C, which uses the parameter of the last
column in Table \ref{tab:vars_and_pars}).

\begin{figure}
\begin{centering}
\includegraphics[angle=-90,width=0.6\paperwidth]{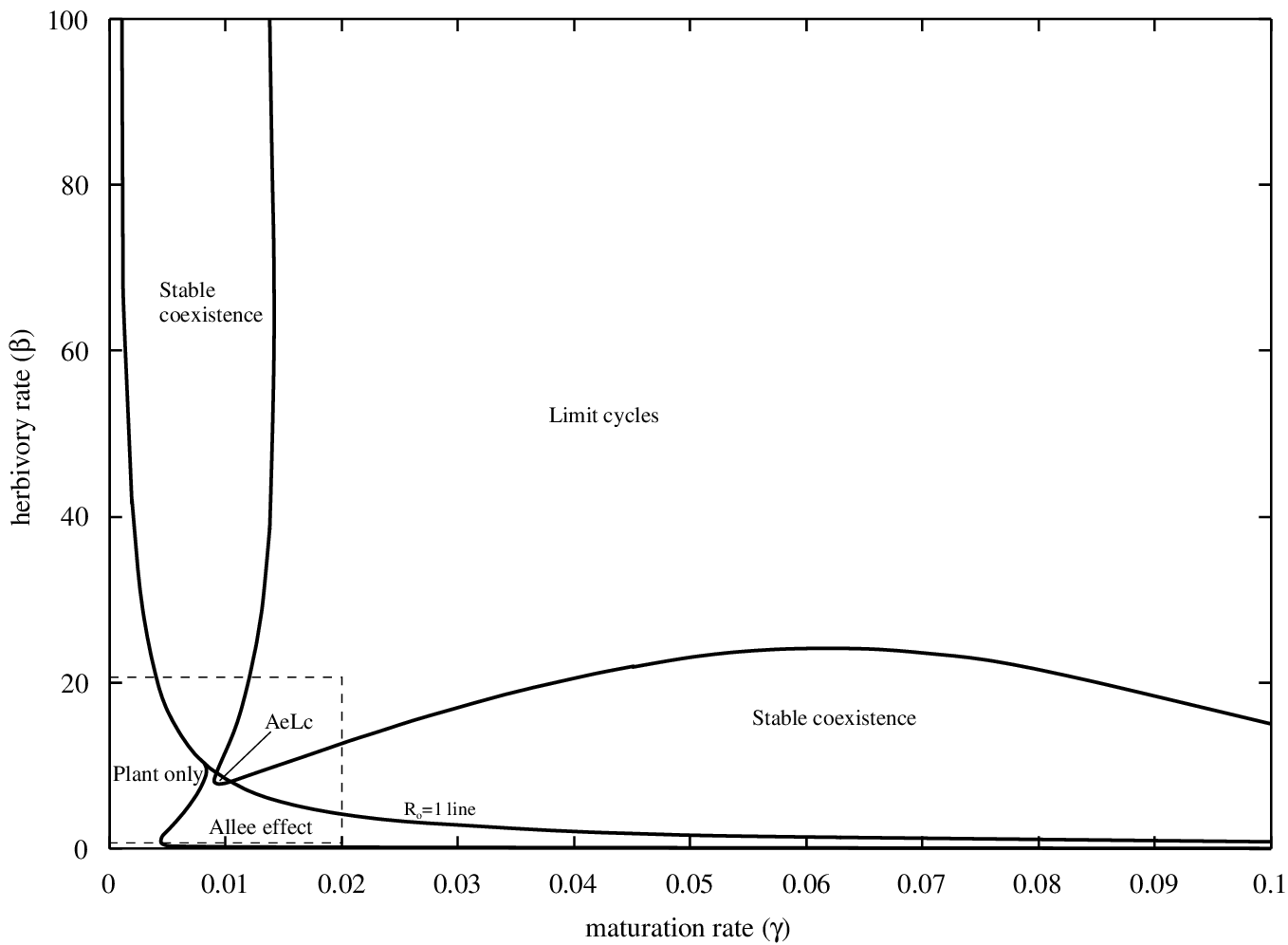}
\par\end{centering}

\protect\caption{\label{fig:beta_vs_gamma_phi1} Outcomes of the PLA model as a function
of the larval maturation and herbivory rates for generalist pollinators
$(\phi=1)$. AeLc: intersection of the Allee effect and Limit cycle
zones.}
\end{figure}

Figure \ref{fig:beta_vs_gamma_phi1} shows the $\beta$ vs $\gamma$
parameter space of the model when the adults are more generalist.
The relative positions of the plant-only, Allee effect, and coexistence
regions are similar to the case of specialist pollinators (Figure
\ref{fig:beta_vs_gamma_phi0}). However, the region of limit cycles
is much larger. The $R_{0}=1$ line is closer to the origin, because
the expression for $R_{0}$ is now (see derivation in Appendix A):

\begin{equation}
R_{0}=\frac{(\epsilon\alpha+\phi\eta)\gamma\beta}{\eta\nu(\mu+\gamma\beta)}\label{eq:r0gen}
\end{equation}

In other words, this means that the more generalist the adult pollinators
(larger $\phi$), the more likely they can invade when rare. There
is also a small overlap between the Allee effect and limit cycle regions,
i.e. parameter combinations for which the long term outcome could
be insect extinction or plant--insect oscillations, depending on the
initial conditions.

\subsection{Graphical analysis\label{sub:analytical}}

The general features of the interaction can be studied by phase-plane
analysis. To make this easier, we collapsed the three-dimensional
PLA model into a two-dimensional plant--larva (PL) model, by assuming
that adults are extremely short lived compared with plants and larvae
(see resulting ODE in Appendix B). The closest realization of this
assumption could be \emph{Manduca sexta}, which has a larval stage
of approximately 20-25 days and adult stages of around 7 days \citep{Reinecke_1980,Ziegler_1991}.
For a given parametrization (Table \ref{tab:vars_and_pars}), the
PL model has the same equilibria as the PLA model, but not the exact
same global dynamics due to the alteration of time scales. Yet, this
simplification provides insights about the outcomes displayed in Figures
\ref{fig:beta_vs_gamma_phi0} and \ref{fig:beta_vs_gamma_phi1}.

\begin{figure}
\begin{centering}
\includegraphics[angle=-90,width=0.75\paperwidth]{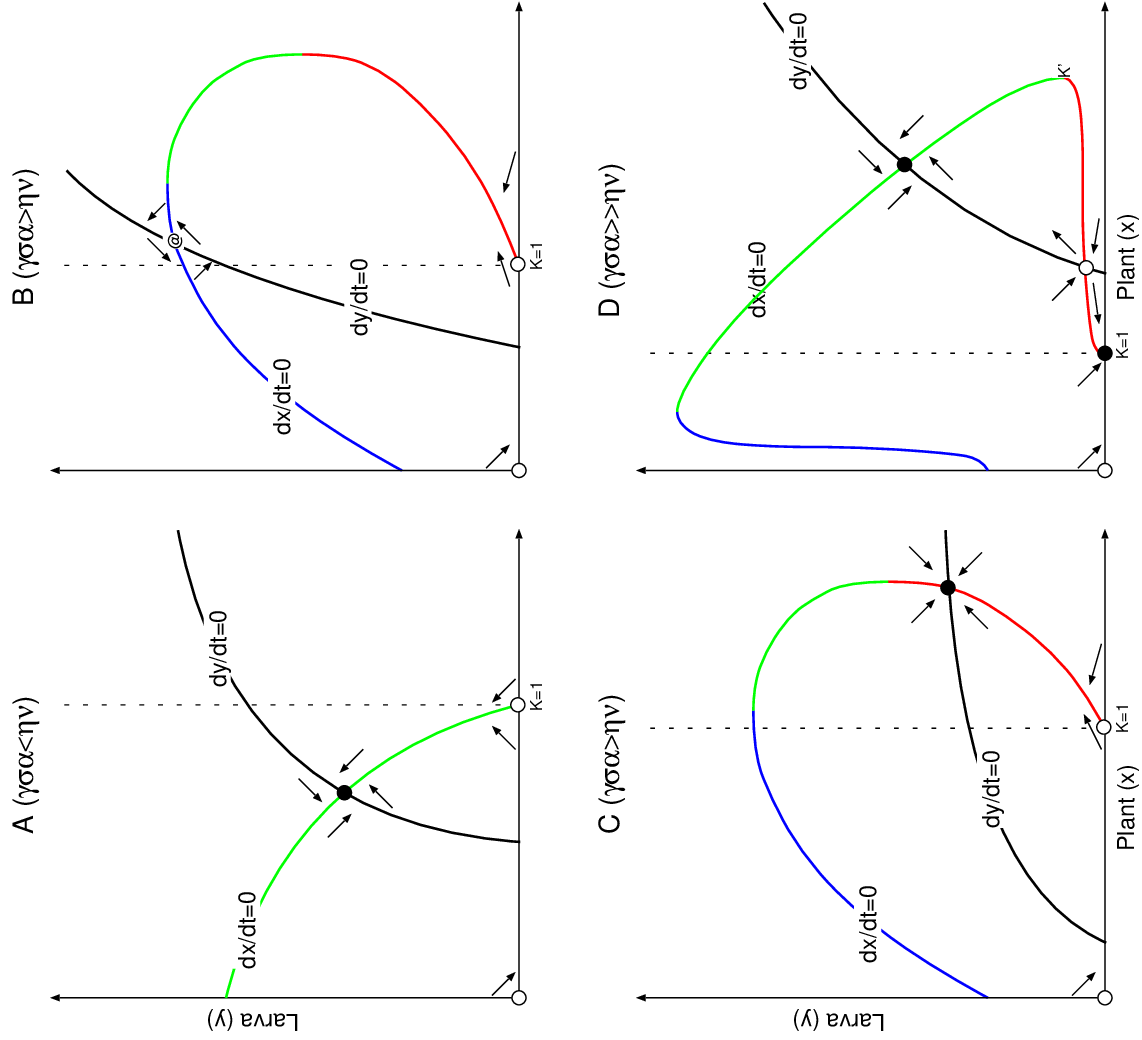}
\par\end{centering}

\protect\caption{\label{fig:phasespace} Isoclines and vector fields in the simplified
version of the PLA model. The vertical line at K separates zones above
and below the plant's carrying capacity $(x=1)$. Equilibria (intersections)
can be stable (black), unstable (white), or unstable focus (@, see
text for justification). (A) An always decreasing plant isocline (green)
intersects the larva isocline (black) leading to stable coexistence
via damped oscillations. (B) Plant isoclines can bulge above K (green+red)
and have a hump (blue+green). Intersections at the left of the hump
(blue) lead to limit cycles, intersections at the right (green) lead
to damped oscillations like in (A). (C) The plant's isocline is like
in (B), but the insect's isocline intersects the ``fold'' (red)
of the plant's isocline, resulting in stable equilibria without oscillations.
(D) A mushroom-shaped plant's isocline whose decreasing part (green)
resembles a logistic prey isocline with enlarged pseudo carrying capacity
(K'>1). Isocline intersections can happen once, or twice as depicted,
giving rise to Allee effects (this can also happen with plant isoclines
like in B,C but never in A).}
\end{figure}

Figure \ref{fig:phasespace} shows plant and larva isoclines (i.e.
non-trivial nullclines) and coexistence equilibria (intersections).
Isocline properties are analytically justified (Appendix B). These
sketches are grossly exaggerated, but this facilitates the representation
of features that are hard to notice by plotting them numerically (e.g.
with parameters like in Table \ref{tab:vars_and_pars}).

Plant isoclines take two main forms:

\begin{equation}
\begin{cases}
\gamma\sigma\alpha<\eta\nu & \textrm{the isocline lies entirely below (to the left of) the carrying capacity}\\
\gamma\sigma\alpha>\eta\nu & \textrm{parts of the isocline lie above (to the right of) the carrying capacity}
\end{cases}\label{eq:pcases}
\end{equation}

\noindent In both cases, plants grow between the isocline and the
axes, and decrease otherwise. Larva isoclines are simpler, they start
in the plant axis and bend towards the right when insects tend towards
specialization ($\phi<\nu$). When insects tend towards generalism
($\phi>\nu$), their isoclines increase rapidly upwards like the letter
``J'' (not shown here, see Appendix B) . Insects grow below and
right of the larva isocline, and decrease otherwise.

The $\gamma\sigma\alpha<\eta\nu$ case in Figure \ref{fig:phasespace}A
covers scenarios in which pollination rates $(\alpha)$, plant benefits
$(\sigma)$, adult pollinator lifetimes $(1/\nu)$ and larva-to-adult
transition rates $(\gamma)$ are low. The plant's isocline is a decreasing
curve crossing the plant's axis at its carrying capacity K $(x=1,y=0)$.
Coexistence is unfavorable for the plant since its equilibrium biomass
lies below the carrying capacity $(x<1)$. The local dynamics around
the coexistence equilibrium indicates oscillations, and we can use
the geometry of the intersection to infer that the equilibrium is
stable (eigenvalue analysis is too difficult to perform for this model):
Figure \ref{fig:phasespace}A shows that if plants increase (or decrease)
above the equilibrium while keeping the insect density fixed, they
enter a zone of negative (or positive) growth; and the same holds
for the insects while keeping the plants fixed. In ecological terms,
both species are self-limited around the coexistence equilibrium,
which as a rule of thumb is a strong indication of stability \citep{case2000}.
Together with the fact that the trivial $(x=0,y=0)$ and carrying
capacity equilibrium $(x=1,y=0)$ are saddle points, we conclude that
plants and insects achieve an equilibrium after a period of transient
oscillations (provided that insects are viable, e.g. $\beta,\gamma,\epsilon$
are large enough). Indeed, for extreme scenarios of negligible plant
pollination benefits (i.e. $\alpha$ and/or $\sigma$ tend to zero),
the plant's isocline approximates a straight line with a negative
slope, like the isocline of a logistic prey in a Lotka--Volterra model,
which is well known to cause damped oscillations \citep{case2000}.

The $\gamma\sigma\alpha>\eta\nu$ case in Figures \ref{fig:phasespace}B,C,D
covers scenarios in which pollination rates $(\alpha)$, pollination
benefits $(\sigma)$, adult pollinator lifetimes $(1/\nu)$ and larva-to-adult
(harm-to-benefit) transition rates $(\gamma)$ are high. One part
of the plant's isocline lies above the carrying capacity, which means
that coexistence equilibria with plant biomass larger than the carrying
capacity $(x>1)$ are possible; this is favorable for the plant. The
isocline also displays a ``hump'' like in the classical victim--exploiter
models \citep{rosenzweig_macarthur-amnat63}. Intersections at the
right of the hump would lead to damped oscillations, like in Figure
\ref{fig:phasespace}A (for $\gamma\sigma\alpha<\eta\nu$). Intersections
at the left of the hump, like in Figure \ref{fig:phasespace}B, suggest
oscillations that will increase in amplitude. This is because a small
increase (decrease) along the plant's axis leaves the plant at the
growing (decreasing) side of its isocline, promoting further increase.
This means that plants do not experience self-limitation, a rule-of-thumb
indicator of instability \citep{case2000} and we infer that interactions
would not dampen, leading to limit cycles. We have seen in Figure
\ref{fig:dynamics_muther} that limit cycles can pass above the plant's
carrying capacity, which is implied in Figure \ref{fig:phasespace}B,
by picturing the maximum of the plant's hump at the right of carrying
capacity, and the intersection of isoclines between both points. Even
if the hump lies at left of the carrying capacity, we cannot use this
graphical analysis to discard the possibility of limit cycles overcoming
the carrying capacity.

For $\gamma\sigma\alpha>\eta\nu$ the plant's isocline also ``folds''
from its rightmost extent back towards the carrying capacity point.
An intersection with this fold is shown in Figure \ref{fig:phasespace}C,
resulting in an equilibrium above the plant's carrying capacity $(x>1)$,
which is approached without oscillations. Intersections can also result
in two equilibria, in which one of them is always unstable and belongs
to a threshold above which insect invasion is possible (and not possible
if below) (Figure \ref{fig:phasespace}D). This explains the Allee
effect, i.e. insect intrinsic growth rates increase from negative
to positive as insect initial density increases.

When the second inequality of (\ref{eq:pcases}) widens $(\gamma\sigma\alpha\gg\eta\nu)$,
the plant's isocline tends to take a mushroom-like shape (or ``anvil''
or letter ``$\Omega$''), as in Figure \ref{fig:phasespace}D. Indeed,
as $\gamma,\sigma,\alpha$ increase and/or $\eta,\nu$ decrease more
and more, the decreasing segment of the isocline approximates a decreasing
line, while the rest of the isocline is pushed closer and closer to
the axes. In other words, when pollination rates $(\alpha)$, benefits
$(\sigma)$, adult lifetimes $(1/\nu)$ and larva development rates
$(\gamma)$ increase, plant isoclines would resemble the isocline
of a logistic prey, with a ``pseudo'' carrying capacity (the rightmost
extent of the isocline) larger than the intrinsic carrying capacity
($x=1$). These conditions would promote stable coexistence with large
plant equilibrium biomasses.

\section{Discussion\label{sec:Discussion}}

We developed a plant--insect model that considers two interaction
types, pollination and herbivory. Ours belongs to a class of models
\citep{hernandez-rspb98,holland_deangelis-ecology10} in which balances
between costs and benefits cause continuous variation in interaction
strengths, as well as transitions among interaction types (mutualism,
predation, competition). In our particular case, interaction types
depend on the stage of the insect's life cycle, as inspired by the
interaction between \emph{M. sexta} and \emph{D. wrightii} \citep{bronstein_etal-annbot09,alarcon_etal-ecoent08}
or between \emph{M. sexta} and \emph{N. attenuata} \citep{baldwin-oecologia88}.
There are many other examples of pollination--herbivory in Lepidopterans,
where adult butterflies pollinate the same plants exploited by their
larvae \citep{wackers_etal-are07,altermatt_pearse-amnat11}. We assign
antagonistic and mutualistic roles to larva and adult insect stages
respectively, which enable us to study the consequences of ontogenetic
changes on the dynamics of plant--insect associations, a topic that
is receiving increased attention \citep{miller_rudolf-tree11,rudolf_lafferty-ecolett11}.
Our model could be generalized to other scenarios, in which drastic
ontogenetic niche shifts cause the separation of benefits and costs
in time and space. But excludes cases like the yucca/yucca moth interaction
\citep{holland_etal-amnat02}, where adult pollinated ovules face
larval predation, i.e. benefits themselves are deducted.

Instead of using species biomasses as resource and service proxies
\citep{holland_deangelis-ecology10}, we consider a mechanism (\ref{eq:pfla})
that treats resources more explicitly \citep{encinas_etal-jtb14}.
We use flowers as a direct proxy of resource availability, by assuming
a uniform volume of nectar per flower. Nectar consumption by insects
is concomitant with service exploitation by the plants (pollination),
based on the assumption that flowers contain uniform numbers of ovules.
Pollination also leads to flower closure \citep{primack-ares85},
making them limiting resources. Flowers are ephemeral compared with
plants and insects, so we consider that they attain a steady-state
between production and disappearance. As a result, the dynamics is
stated only in terms of plant, larva and adult populations, i.e. the
PLA model (\ref{eq:pla}). The feasibility of the results described
by our analysis depends on several parameters. The consumption, mortalities
and growth rates, and the carrying capacities (e.g. $a,b,m,n$ and
$r,c$ in the fourth column of Table \ref{tab:vars_and_pars}), have
values close to the ranges considered by other models \citep{holland_deangelis-ecology10,johnson_amarasekare-jtb13}.
Oscillations, for example, require large herbivory rates, but this
is usual for \emph{M. sexta} \citep{McFadden_1968}.

\subsection{Mutualism--antagonism cycles}

The PLA model displays plant--insect coexistence for any combination
of (non-trivial) initial conditions where insects can invade when
rare $(R_{o}>1)$. Coexistence is also possible where insects cannot
invade when rare $(R_{o}<1)$, but this requires high initial biomasses
of plants and insects (Allee effect). Coexistence can take the form
of a stable equilibrium, but it can also take the form of stable oscillations,
i.e. limit cycles.

Previous models combining mutualism and antagonism predict oscillations,
but they are transient ones \citep{holland_etal-amnat02,wang_deangelis-mbe12},
or the limit cycles occur entirely below the plant's carrying capacity
\citep{holland_etal-theorecol13}. We have good reasons to conclude
that the cycles are herbivory driven and not simply a consequence
of the PLA model having many variables and non-linearities. First
of all, limit cycles require herbivory rates $(\beta)$ to be large
enough. Second, given limit cycles, an increase in the maturation
rate $(\gamma)$ causes a transition to stable coexistence, and further
increase in $\beta$ is required to induce limit cycles again (Figure
\ref{fig:beta_vs_gamma_phi0}). This makes sense because by speeding
up the transition from larva to adult, the total effect of herbivory
on the plants is reduced, hence preventing a crash in plant biomass
followed by a crash in the insects. Third, when adult pollinators
have alternative food sources $(\phi>1)$, the zone of limit cycles
in the space of parameters becomes larger (Figure \ref{fig:beta_vs_gamma_phi1}).
This also makes sense, because the total effect of herbivory increases
by an additional supply of larva (which is not limited by the nectar
of the plant considered), leading to a plant biomass crash followed
by insect decline.

The graphical analysis provides another indication that oscillations
are herbivory driven. On the one hand insect isoclines (or rather
larva isoclines) are always positively sloped, and insects only grow
when plant biomass is large enough (how large depends on insect's
population size, due to intra-specific competition). Plant isoclines,
on the other hand, can display a hump (Figure \ref{fig:phasespace}B,C,D),
and they grow (decrease) below (above) the hump. These two features
of insect and plant isoclines are associated with limit cycles in
classical victim--exploiter models (\citealt{rosenzweig_macarthur-amnat63}).
If there is no herbivory or another form of antagonism (e.g. competition)
but only mutualism, the plant's isocline would be a positively sloped
line, and plants would attain large populations in the presence of
large insect populations, without cycles. However, mutualism is still
essential for limit cycles: if mutualistic benefits are not large
enough $(\gamma\sigma\alpha<\eta\nu)$, plant isoclines do not have
a hump (Figure \ref{fig:phasespace}A) and oscillations are predicted
to vanish. The effect of mutualism on stability is like the effect
of enrichment on the stability in pure victim--exploiter models \citep{rosenzweig-science71},
by allowing the plants to overcome the limits imposed by their intrinsic
carrying capacity (e.g. the pseudo-carrying capacity K' in Figure
\ref{fig:phasespace}D).

\subsection{Classification of outcomes: mutualism or herbivory?}

Interactions can be classified according to the net effect of one
species on the abundance (biomass, density) of another (but see other
schemes \citealt{abrams-oecologia87}). This classification scheme
can be problematic in empirical contexts, because reference baselines
such as carrying capacities are usually not known and because stable
abundances make little sense under the influence of unpredictable
external fluctuations \citep{hernandez-jtb09}.

Our PLA model illustrates the classification issue when non-equilibrium
dynamics are generated endogenously, i.e. not by external perturbations.
Since plants are facultative mutualists and insects are obligatory
ones, one can say the outcome is \emph{net mutualism} (+,+) or \emph{net
herbivory} (--,+), if the coexistence is stable, and the plant equilibrium
ends up respectively above or below the carrying capacity \citep{hernandez-rspb98,holland_deangelis-ecology10}.
If coexistence is under non-equilibrium conditions and plant oscillations
are entirely below the carrying capacity (e.g. for large $\beta$),
the outcome is detrimental for plant and hence there is net herbivory
(--,+); oscillations may in fact be considered irrelevant for this
conclusion (or may further support the case of herbivory, read below).
However, when the plant oscillation maximum is above carrying capacity
and the minimum is below, like in Figure \ref{fig:dynamics_muther},
could we say that the system alternates periodically between states
of net mutualism and net herbivory? Here perhaps a time-based average
over the cycle can help up us decide. The situation could be more
complicated if plant oscillations lie entirely above the carrying
capacity (see an example in Appendix C): one can say that the net
outcome is a mutualism due to enlarged plant biomasses, but the oscillations
indicates that a victim--exploiter interaction exists. As we can see,
deciding upon the net outcome require consideration of both equilibrium
and dynamical aspects.

\subsection{Factors that could cause dynamical transitions}

\subsubsection*{Environmental factors}

The parameters in our analyses can change due to external factors.
One of the most important is temperature \citep{gillooly_etal-science01}.
It is well known for example, that warming can reduce the number of
days needed by larvae to complete their development \citep{bonhomme-ejagr00},
making $\gamma$ higher. Keeping everything else equal but $\gamma$,
for insects that cannot invade when rare (i.e. displaying Allee effects,
$R_{o}<1$), a cooling of the environment will cause the sudden extinction
of the insect and a catastrophic collapse of the mutualism, which
cannot be simply reverted by warming. For insects that can invade
when rare $(R_{o}>1)$, by slowing down larva development, cooling
would increase the burden of herbivory over the benefits of pollination
making the system more prone to oscillations and less stable (even
less under strong herbivory, large $\beta$). Flowering, pollination,
herbivory, growth and mortality rates (e.g. $s,a,b,r,m$ and $n$
in equations \ref{eq:pfla}) are also temperature-dependent, and they
can increase or decrease with warming depending on the thermal impacts
on insect and plant metabolisms \citep{vasseur_mccann-amnat05}. This
makes general predictions more difficult. However, we get the general
picture that warming or cooling can change the balance between costs
and benefits impacting the stability of the plant--insect association.

Dynamical transitions can also be induced by changes in the chemical
environment, often as a consequence of human activity. Some pesticides,
for example, are hormone retarding agents \citep{dev-pinsa86}. This
means that their release can reduce $\gamma$ altering the balance
of the interaction towards more herbivory and less pollination and
finally endangering pollination service \citep{Potts_2010_tree,Kearns_1998}.
In other cases, the chemical changes are initiated by the plants:
in response to herbivory, many plants release predator attractants
\citep{allmann_baldwin-science10}, which can increase larval mortality
$\mu$. If the insect does nothing but harm, this is always an advantage.
If the insect is also a very effective pollinator, the abuse of this
strategy can cost the plant important pollination services because
a dead herbivore today is one less pollinator tomorrow.

Another factor that can increase or decrease larvae maturation rates
($\gamma$), is the level of nutrients present in the plant's vegetative
tissue \citep{Woods_1999,Perkins_2004}. On the one hand, the use
of fertilizers rich in phosphorus could increase larvae maturation
rates \citep{Perkins_2004}. On the other hand, under low protein
consumption \emph{M. sexta} larvae could decrease maturation rate,
although \emph{M. sexta} larvae can compensate this lack of proteins
by increasing their herbivory levels (i.e. compensatory consumption)
\citep{Woods_1999}. Thus, different external factors related to plant
nutrients could indirectly trigger different larvae maturation rates
that will potentially modify the interaction dynamics.

\subsubsection*{Pollinator's diet breadth}

An important factor that can affect the balance between mutualism
and herbivory is the diet breadth of pollinators. Alternative food
sources for the adults could lead to apparent competition \citep{holt-tpb77}
mediated by pollination, as predicted for the interaction between
\emph{D. wrigthii} (Solanacea) and \emph{M. sexta} (Sphingidae) in
the presence of \emph{Agave palmieri} (plant) \citep{bronstein_etal-annbot09}:
visitation of \emph{Agave} by \emph{M. sexta} does not affect the
pollination benefits received by \emph{D. wrightii}, but it increases
oviposition rates on \emph{D}. \emph{wrightii}, increasing herbivory.
As discussed before, such an increase in herbivory could explain why
oscillations are more widespread when adult insects have alternative
food sources $(\phi>0)$ in our PLA model.

Although we did not explore this with our model, the diet breadth
of the larva could also have important consequences. In the empirical
systems that inspired our model, the larva can have alternative hosts
\citep{alarcon_etal-ecoent08}, spreading the costs of herbivory over
several species. The local extinction of such hosts could increase
herbivory on the remaining ones, promoting unstable dynamics. To explore
these issues properly, models like ours must be extended to consider
larger community modules or networks, taking into account that there
is a positive correlation between the diet breadths of larval and
adult stages \citep{altermatt_pearse-amnat11}.

From the perspective of the plant, the lack of alternative pollinators
could also lead to increased herbivory and loss of stability. The
case of the tobacco plant (\emph{N. attenuata}) and \emph{M. sexta}
is illustrative. These moths are nocturnal pollinators, and in response
to herbivory by their larvae, the plants can change their phenology
by opening flowers during the morning instead. Thus, oviposition and
subsequent herbivory can be avoided, whereas pollination can still
be performed by hummingbirds \citep{kessler_etal-cb10}. Although
hummingbirds are thought to be less reliable pollinators than moths
for several reasons \citep{irwin-cb10}, they are an alternative with
negligible costs. Thus, a decline in hummingbird populations will
render the herbivore avoidance strategy useless and plants would have
no alternative but to be pollinated by insects with herbivorous larvae
that promote oscillations.

\subsection{Conclusions}

Many insect pollinators are herbivores during their larval phases.
If pollination and herbivory targets the same plant (e.g. as between
tobacco plants and hawkmoths), the overall outcome of the association
depends on the balance between costs and benefits for the plant. As
predicted by our plant-larva-adult (PLA) model, this balance is affected
by changes in insect development: the faster larvae turns into adults
the better for the plant, and the more stable the interaction; the
slower this development the poorer the outcome for the plant, and
the less stable the interaction (oscillations). Under plant--insect
oscillations, this balance can be dynamically complex (e.g. periodic
alternation between mutualism and antagonism). Since maturation rates
play an essential role in long term stability, we predict important
qualitative changes in the dynamics due to changes in environmental
conditions, such as temperature and chemical compounds (e.g. toxins,
hormones, plant nutrients). The stability of these mixed interactions
can also be greatly affected by changes in the diet generalism of
the pollinators.

\section*{Acknowledgements}

We thank Rampal Etienne for the discussions that inspired us to write
this article. We thank the comments and suggestions from our colleagues
of the Centre for Biodiversity Theory and Modelling in Moulis, France,
and the Centre for Australian National Biodiversity and Research at
CSIRO in Canberra, Australia. TAR was supported by the TULIP Laboratory
of Excellence (ANR-10-LABX-41). FEV was supported by the OCE postdoctoral
fellowship at CSIRO.

\noindent \begin{flushright}

\par\end{flushright}

\bibliographystyle{chicago}
\bibliography{plamodel}

\appendix

\section*{Appendices}

\addcontentsline{toc}{section}{Appendices}

\subsection*{Appendix A: Bifurcations}

\addcontentsline{toc}{subsection}{Appendix A}
\renewcommand{\theequation}{A.\arabic{equation}} \setcounter{equation}{0}
\renewcommand{\thefigure}{A.\arabic{figure}} \setcounter{figure}{0}

\begin{figure}
\begin{centering}
\includegraphics[clip,angle=-90,scale=0.5]{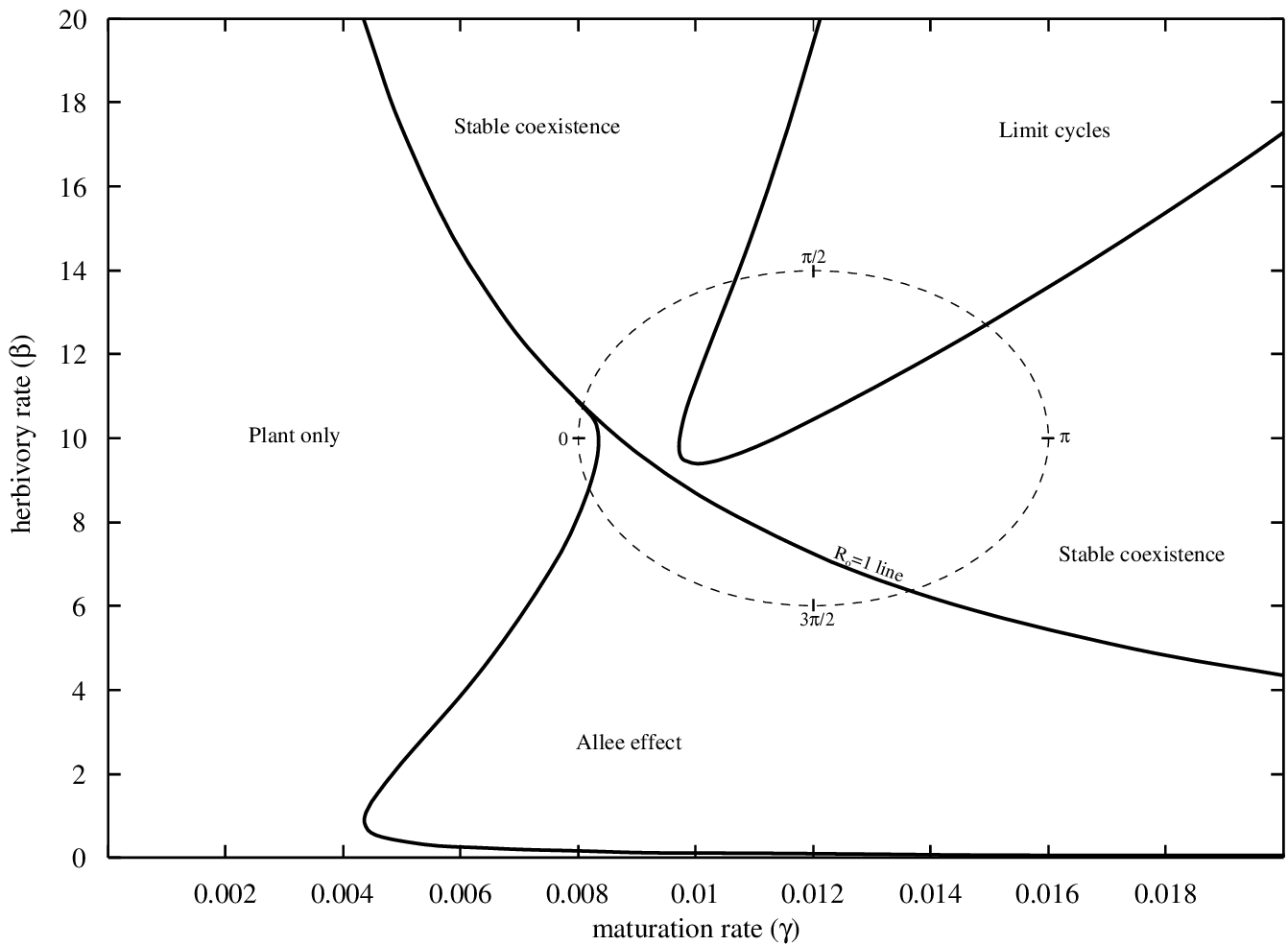}
\par\end{centering}

\protect\caption{\label{fig:app.beta_vs_gamma_phi0}Outcomes of the PLA model as a
function of the larval maturation and herbivory rates for specialist
pollinators. The ellipse describes the joint variation of $\gamma$
and $\beta$ taking place in the bifurcation diagram in Figure \ref{fig:app.x_vs_angle_phi0}.}
\end{figure}

Figure \ref{fig:beta_vs_gamma_phi0} in the main text shows all outcomes
(plant-only, Allee effect, stable coexistence and limit cycles) occurring
together in a rectangle at the bottom left corner of the parameter
space $\beta$ vs $\gamma$. We enlarged this rectangle in Figure
\ref{fig:app.beta_vs_gamma_phi0} in order to show the bifurcations
of the PLA model as we traverse the parameter space along an elliptical
path as indicated.

\begin{figure}
\begin{centering}
\includegraphics[clip,angle=-90,scale=0.5]{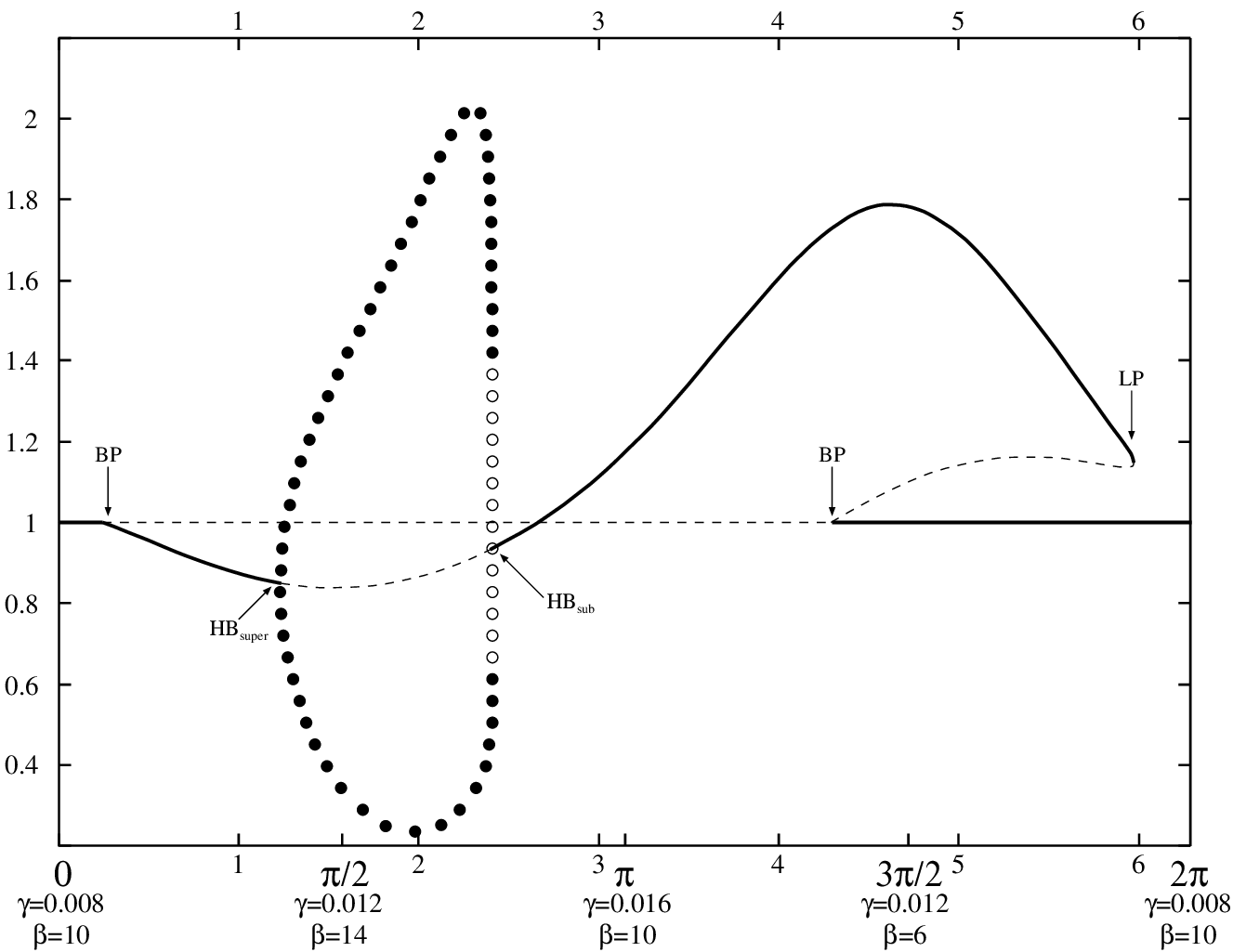}
\par\end{centering}

\protect\caption{\label{fig:app.x_vs_angle_phi0}Bifurcation diagram along the elliptical
path drawn in Figure \ref{fig:app.beta_vs_gamma_phi0}, with reference
values of $\beta$ and $\gamma$ for each quarter of a rotation. Solid
(broken) lines represent stable (unstable) equilibria, black (white)
circles represent limit cycle maxima and minima. The $x=1$ line corresponds
to the plant carrying capacity. HB\protect\textsubscript{super}:
super-critical and HB\protect\textsubscript{sub}: sub-critical Hopf
bifurcations, BP: branching point (transcritical bifurcation), LP:
limit point (fold bifurcation).}
\end{figure}

From Figure \ref{fig:app.x_vs_angle_phi0} we can conclude that plant
equilibrium biomasses (stable or not) are inversely related with the
rate of herbivory $(\beta)$. A similar response occurs regarding
oscillations: as long as $\beta$ values are large enough to induce
oscillations (the part in the figure marked with circles), such oscillations
tend to display lower maxima and minima for larger values of $\beta$,
and higher maxima and minima for smaller values instead.

The response of plant biomasses with respect to the insect maturation
rate $(\gamma)$ is more complex. For example around the middle part
of Figure \ref{fig:app.x_vs_angle_phi0} (between the $\pi/2$ and
$3\pi/2$ marks), increasing $\gamma$ causes (equilibrium) plant
biomass increases if herbivory is high, but decreases if herbivory
is low. In contrast, increasing $\gamma$ from very low values causes
plant biomass to increase if herbivory is low (between LP and the
$3\pi/2$ mark at the right) or decrease when it is high (between
BP and the $\pi/2$ mark at the left).

The transitions between stability and limit cycles are typically \emph{super-critical
Hopf bifurcations}, in which a stable branch of periodic solutions
overlaps a branch of unstable equilibria. The bifurcation diagram
(Figure \ref{fig:app.x_vs_angle_phi0}) also displays a \emph{sub-critical
Hopf bifurcation}, in which an unstable branch of periodic solutions
overlaps stable equilibria. In such cases the long term outcome can
be stable coexistence or a limit cycle depending on the initial conditions.
Given the parameter values in Table \ref{tab:vars_and_pars}, this
sub-critical Hopf bifurcation zone was too narrow to be represented
in the parameter space (Figure \ref{fig:app.beta_vs_gamma_phi0}).
Appendix C contains a simulation in which a small change in the initial
conditions causes the system to approach an equilibrium or a limit
cycle.

The $R_{o}=1$ line in Figure \ref{fig:beta_vs_gamma_phi0} can be
found analytically. To do this, we need to know when the carrying
capacity equilibrium switches between stable and unstable, which depends
on the eigenvalues of the jacobian matrix of the PLA model (\ref{eq:pla})
evaluated at $(x,y,z)=(1,0,0)$:

\begin{equation}
\left[\begin{array}{ccc}
1-2x+\frac{\sigma\alpha z}{\eta+z}-\beta y & -\beta x & \frac{\sigma\alpha\eta x}{(\eta+z)^{2}}\\
\frac{\epsilon\alpha z}{\eta+z}-\gamma\beta y & -\mu-\gamma\beta x & \frac{\epsilon\alpha\eta x}{(\eta+z)^{2}}+\phi\\
\gamma\beta y & \gamma\beta x & -\nu
\end{array}\right]=\left[\begin{array}{ccc}
-1 & -\beta & \frac{\sigma\alpha}{\eta}\\
0 & -\mu-\gamma\beta & \frac{\epsilon\alpha}{\eta}+\phi\\
0 & \gamma\beta & -\nu
\end{array}\right]\label{eq:jacobian}
\end{equation}

The eigenvalues of the jacobian are $\lambda_{1}=-1$ and:

\[
\lambda_{2}=\frac{-(\mu+\nu+\gamma\beta)\pm\sqrt{(\mu+\nu+\gamma\beta)^{2}-4\left[\nu(\mu+\gamma\beta)-\gamma\beta(\phi+\epsilon\alpha/\eta)\right]}}{2}
\]

\noindent thus $(x,y,z)=(1,0,0)$ is unstable if at least one of $\lambda_{2}$
have a positive real part. This can only happen when:

\begin{equation}
\frac{(\epsilon\alpha+\phi\eta)\gamma\beta}{\eta\nu(\mu+\gamma\beta)}>1\label{eq:invasion}
\end{equation}

\noindent by which automatically both $\lambda_{2}$ are real (one
is negative and the other is positive). The right-hand side of (\ref{eq:invasion})
is $R_{o}$ in the main text. Making $R_{o}=1$ and writing $\beta$
as a function of $\gamma$, we obtain a decreasing hyperbolic line
with asymptotes $\beta=0$ and $\gamma=0$ as shown in Figures \ref{fig:beta_vs_gamma_phi0}
and \ref{fig:beta_vs_gamma_phi1}. This is yet another reason, a pure
technical one this time, that explains why we choose to present our
results in the form of a $\beta$ vs $\gamma$ parameter space.

Since the eigenvector of $\lambda_{1}$ is a multiple of $(1,0,0)$,
the eigenvectors of $\lambda_{2}$ are orthogonal to $(1,0,0)$, i.e.
$v=(0,v_{y},v_{z}),w=(0,w_{y},w_{z})$. This, and the fact that both
$\lambda_{2}$ are real if the inequality above holds, means that
only perturbations in $y$ and/or $z$, i.e. an insect invasion, would
make $(x,y,z)=(1,0,0)$ unstable.

\subsection*{Appendix B: Isocline properties}

\addcontentsline{toc}{subsection}{Appendix B}
\renewcommand{\theequation}{B.\arabic{equation}} \setcounter{equation}{0}
\renewcommand{\thefigure}{B.\arabic{figure}} \setcounter{figure}{0}

Let us assume that the adult phase is very short lived compared with
the larval phase and with the dynamics of the plant. In the same way
as we did in the case of the flowers, assume that the adults reach
a steady-state $dz/dt\approx0$ with respect to the other variables,
and that the adult biomass can be approximated by $z\approx\gamma\beta xy/\nu$.
Substituting this in the ODE system (\ref{eq:pla}), we obtain the
two-dimensional system:

\begin{align}
\dot{x} & =x(1-x)+\sigma\frac{\alpha\gamma\beta x^{2}y}{\eta\nu+\gamma\beta xy}-\beta xy\nonumber \\
\dot{y} & =\epsilon\frac{\alpha\gamma\beta x^{2}y}{\eta\nu+\gamma\beta xy}+\frac{\phi\gamma\beta xy}{\nu}-\gamma\beta xy-\mu y\label{eq:pl}
\end{align}

This system has two trivial isoclines, $x=0$ for the plant and $y=0$
for the insect. The following results only concern the non-trivial
isoclines for plants and insects.

\subsubsection*{Plant isocline}

Making $\dot{x}=0$ in (\ref{eq:pl}), the (non-trivial) isocline
of the plant can be written as a polynomial in $x$ and $y$:

\begin{equation}
x^{2}y+\beta xy^{2}-(1+\sigma\alpha)xy+\frac{\eta\nu}{\gamma\beta}x+\frac{\eta\nu}{\gamma}y-\frac{\eta\nu}{\gamma\beta}=0\label{eq:pisocline}
\end{equation}

To characterize the shape of (\ref{eq:pisocline}) we start by finding
asymptotes. To do this we can rewrite (\ref{eq:pisocline}) as a function
of $x$:

\begin{equation}
y(x)=\frac{1}{2\beta}\left\{ \frac{-\left(\frac{\eta\nu}{\gamma}-(1+\sigma\alpha)x+x^{2}\right)\pm\sqrt{\left(\frac{\eta\nu}{\gamma}-(1+\sigma\alpha)x+x^{2}\right)^{2}+4\frac{\eta\nu}{\gamma}x(1-x)}}{x}\right\} \label{eq:fisocline(x)}
\end{equation}

We divide the numerator and the denominator of (\ref{eq:fisocline(x)})
by $x$:

\begin{align*}
y(x) & =\frac{1}{2\beta}\left\{ -\frac{\eta\nu}{\gamma x}+(1+\sigma\alpha)-x\pm\sqrt{\frac{1}{x^{2}}\left(\frac{\eta\nu}{\gamma}-(1+\sigma\alpha)x+x^{2}\right)^{2}+\frac{1}{x^{2}}4\frac{\eta\nu}{\gamma}x(1-x)}\right\} \\
 & =\frac{1}{2\beta}\left\{ -\frac{\eta\nu}{\gamma x}+(1+\sigma\alpha)-x\pm\sqrt{\left(\frac{\eta\nu}{\gamma x}-(1+\sigma\alpha)+x\right)^{2}+4\frac{\eta\nu}{\gamma}\left(\frac{1}{x}-1\right)}\right\} 
\end{align*}

\noindent and we take the limit when $x$ goes to plus or minus infinity:

\begin{align*}
\lim_{x\to\pm\infty}y(x) & =\frac{1}{2\beta}\lim_{x\to\pm\infty}\left\{ 0+(1+\sigma\alpha)-x\pm\sqrt{(0-(1+\sigma\alpha)+x)^{2}+4\frac{\eta\nu}{\gamma}(0-1)}\right\} \\
 & =\frac{1}{2\beta}\lim_{x\to\pm\infty}\left\{ -(x-1-\sigma\alpha)\pm\sqrt{(x-1-\sigma\alpha)^{2}-4\frac{\eta\nu}{\gamma}}\right\} 
\end{align*}

Note that $|x-1-\sigma\alpha|>\sqrt{(x-1-\sigma\alpha)^{2}-4\frac{\eta\nu}{\gamma}}$.
Thus, the square root above can be approximated by $\delta(x)(x-1-\sigma\alpha)$,
where $\delta$ is a number between 0 and 1, and $\delta(x)\to1$
as $x\to\pm\infty$. We can continue as follows:

\begin{equation}
\begin{aligned}\lim_{x\to\pm\infty}y(x) & =\frac{1}{2\beta}\lim_{x\to\pm\infty}\{-(x-1-\sigma\alpha)\pm\delta(x)(x-1-\sigma\alpha)\}\\
 & =\frac{x-1-\sigma\alpha}{\beta}\lim_{x\to\pm\infty}\frac{\{-1\pm\delta(x)\}}{2}
\end{aligned}
\label{eq:slantlimit}
\end{equation}

When $x\to\pm\infty$ and $\delta\to1$, the '+' branch, $y(x)$ approaches
the horizontal asymptote $y=0$. For this '+' branch we also have
that $-1<\{-1+\delta(x)\}<0$ in (\ref{eq:slantlimit}), which means
that $y$ is negative when $x\to+\infty$, and positive when $x\to-\infty$.
In other words, the horizontal asymptote is approached from below
when $x\to+\infty$ and from above when $x\to-\infty$.

When $x\to\pm\infty$ and $\delta\to1$, the '--' branch, $y(x)$
approaches the slanted asymptote:

\begin{equation}
y=\frac{1+\sigma\alpha-x}{\beta}\label{eq:pasymptote}
\end{equation}

\noindent which decreases with $x$. For this '--' branch we also
have that $-1<\{-1-\delta(x)\}/2<-1/2$ in (\ref{eq:slantlimit}),
which means that when $x\to+\infty$, $y<0$ and $|y|<|(x-1-\sigma\alpha)/\beta|$.
In other words, $y$ lies between 0 and the slanted asymptote when
$x\to+\infty$.

If we write (\ref{eq:pisocline}) as a function of $y$ rather than
as a function of $x$, we will find a vertical asymptote $x=0$, and
the slanted asymptote (\ref{eq:pasymptote}) again. Because (\ref{eq:pisocline})
is symmetric regarding the signs of its terms, the properties of the
vertical asymptote must consistent with those of the horizontal: $y(x)$
goes towards $+\infty$ when $x=0$ is approached from the left, and
towards $-\infty$ when $x=0$ is approached from the right. Also
because of symmetry $x$ must lie between 0 and the slanted asymptote
when $y\to+\infty$.

The following statements tells us the location of special points of
(\ref{eq:pisocline}) as well regions in which (\ref{eq:pisocline})
cannot be satisfied.

\textbf{Lemma 1:} the plant isocline contains the following $(x,y)$
points:

\begin{equation}
\begin{aligned}\mathrm{K} & =(1,0)\\
\mathrm{O} & =(0,\beta^{-1})\\
\mathrm{P} & =(\sigma\alpha-\eta\nu\gamma^{-1},\beta^{-1})\\
\mathrm{Q} & =(1,(\sigma\alpha-\eta\nu\gamma^{-1})\beta^{-1})
\end{aligned}
\label{eq:KOPQ}
\end{equation}

\emph{Proof:} evaluate (\ref{eq:pisocline}) at $x=1$ to get a quadratic
equation in $y$ with roots $y=0$ and $y=(\sigma\alpha-\eta\nu/\gamma)/\beta$,
this gives points K and Q respectively. Evaluate (\ref{eq:pisocline})
at $y=\beta^{-1}$ to get a quadratic equation in $x$ with roots
$x=0$ and $x=\sigma\alpha-\eta\nu/\gamma$, this gives points O and
P respectively. Points K (the plant's carrying capacity), and O are
always biologically feasible (both have non-negative coordinates).

\textbf{Corollary 1:} Simple observation of (\ref{eq:KOPQ}) tells
us that points P and Q are simultaneously biologically feasible if
$\gamma\sigma\alpha>\eta\nu$. Conversely, both are unfeasible if
$\gamma\sigma\alpha<\eta\nu$.

\textbf{Lemma 2:} Points P and Q lie below the slanted asymptote (\ref{eq:pasymptote}).

\emph{Proof:} substitute $y=\beta^{-1}$ in (\ref{eq:pasymptote})
to obtain point $(\sigma\alpha,\beta^{-1})$, and substitute $x=1$
in (\ref{eq:pasymptote}) to obtain point $(1,\sigma\alpha/\beta)$.
Simple inspection shows that point $(\sigma\alpha,\beta^{-1})$ is
always to the right of point P, and point $(1,\sigma\alpha/\beta)$
is always above point Q.

\textbf{Lemma 3:} the plant isocline crosses the x- and y-axis only
at points K and O respectively, and nowhere else.

\emph{Proof:} substituting $y=0$ in (\ref{eq:pisocline}) gives only
one root $x=1$ (i.e. point K). Substituting $x=0$ in (\ref{eq:pisocline})
gives only one root $y=\beta^{-1}$ (i.e. point O).

\textbf{Lemma 4:} the plant isocline is not satisfied in the $(-,-)$
quadrant.

\emph{Proof:} let $a,b\geq0$ and substitute $x=-a$ and $y=-b$ in
(\ref{eq:pisocline}). This leads to:

\begin{equation}
-\left[a^{2}b+\beta ab^{2}+(1+\sigma\alpha)ab+\frac{\eta\nu}{\gamma\beta}a+\frac{\eta\nu}{\gamma}b+\frac{\eta\nu}{\gamma\beta}\right]=0\label{eq:not_in_negneg}
\end{equation}

\noindent since all parameter values are positive, the statement above
is false, thus (\ref{eq:pisocline}) is not satisfied in the $(-,-)$
quadrant.

Using this information about the asymptotes ($x=0,y=0$ and eq. \ref{eq:pasymptote}),
and Lemmas 1, 2, 3 and 4 we can conclude that the plant's isocline
must have one of the two forms depicted in figure \ref{fig:app.piso}.
Corollary 1 explains the form taken in figure \ref{fig:app.piso}A,
when $\gamma\sigma\alpha<\eta\nu$, and the form in figure \ref{fig:app.piso}B,
when $\gamma\sigma\alpha>\eta\nu$. These are the two main cases referenced
in the main text by (\ref{eq:pcases}), where only the positive quadrant
is considered. For points between the O--K segment and the axes $\dot{x}>0$,
otherwise $\dot{x}<0$.

\begin{center}
\begin{figure}
\begin{centering}
\includegraphics[angle=-90,width=0.75\paperwidth]{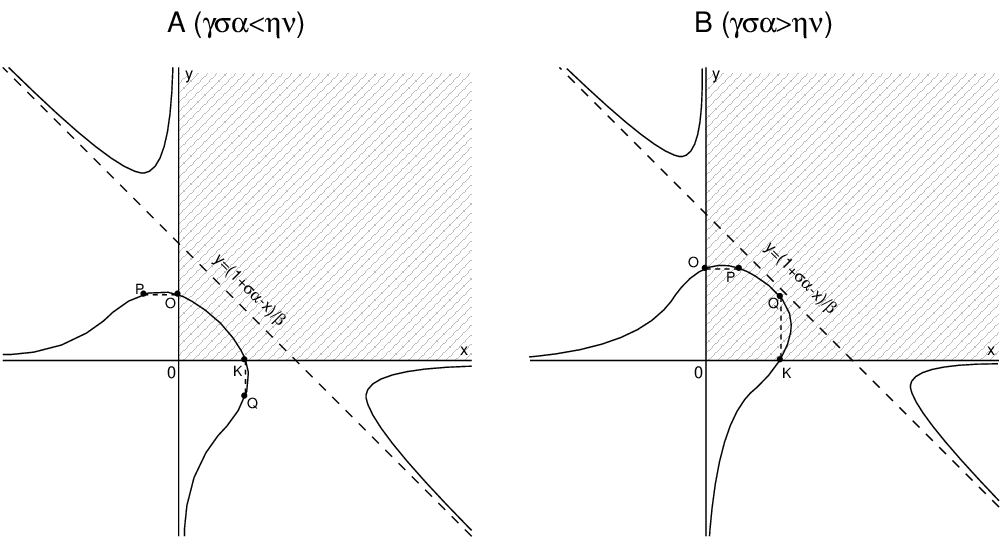}
\par\end{centering}

\protect\caption{\label{fig:app.piso}The two main configurations of the plant isocline.
We only consider the O--K segment in the positive octant (hatched
square). In A the isocline lies below the plant's carrying capacity
(i.e. left of K), in B parts of the isocline lie above (i.e. right
of K).}
\end{figure}

\par\end{center}

Figure \ref{fig:app.pisoshapes} shows how the positive part of the
plant isocline changes as we vary some of the bifurcation parameters.
Increasing $\gamma$ or decreasing $\eta$ or $\nu$, causes the isocline
to be ``compressed'' against the asymptote (\ref{eq:pasymptote})
and it adopts the shape of a mushroom, the letter $\Omega$ or an
anvil. Increasing $\beta$ causes points P and Q to decrease along
the vertically axis. It is more difficult to follow the effect of
the rest of the parameters, for example increasing $\sigma$ and $\alpha$
cause P and Q to move right and upwards respectively, but they also
move the asymptote (\ref{eq:pasymptote}) right and upwards, so we
cannot tell if this will cause the isocline to adopt a mushroom shape.

\begin{center}
\begin{figure}
\begin{centering}
\includegraphics[width=0.75\paperwidth]{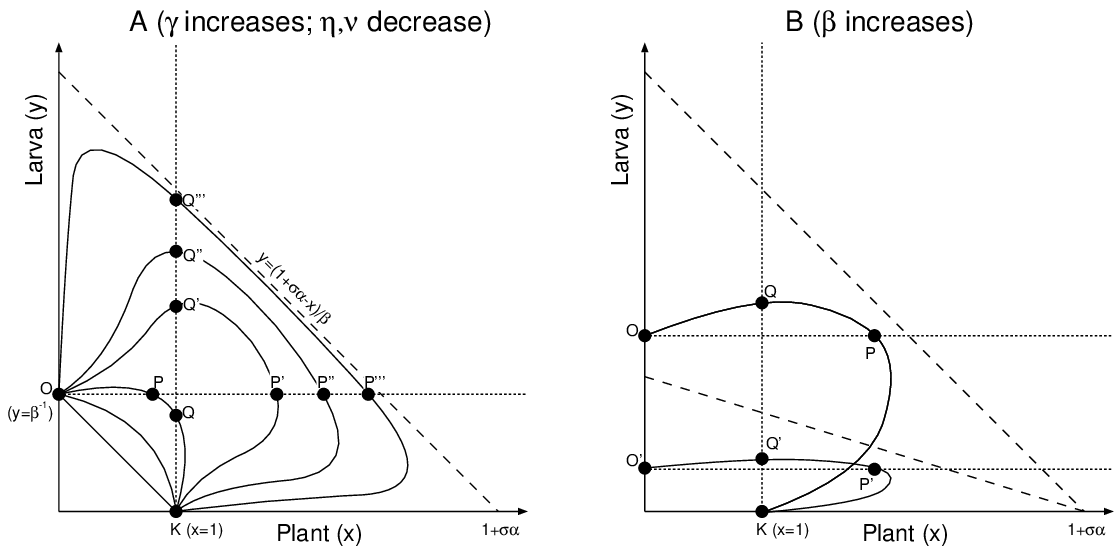}
\par\end{centering}

\protect\caption{\label{fig:app.pisoshapes}Changes in the shape of the plant's isocline.
(A) As $\gamma$ increases and $\eta,\nu$ decrease, points P and
Q move closer to the diagonal asymptote (broken line), and the isocline
eventually adopts the form of a mushroom. (B) As $\beta$ increases,
O, P, Q and the diagonal asymptote move towards the plant axis and
the isocline is compressed vertically.}
\end{figure}

\par\end{center}

\subsubsection*{Larva isocline}

Making $\dot{y}=0$ in (\ref{eq:pl}) the larva isocline is:

\begin{equation}
y(x)=\frac{p(x)}{q(x)}\label{eq:lisocline}
\end{equation}

\noindent where the numerator and denominator:

\begin{align}
p(x) & =\epsilon\alpha\gamma\beta x^{2}-\eta\nu\gamma\beta(1-\phi/\nu)x-\eta\mu\nu\label{eq:polyp}\\
q(x) & =\gamma\beta\left[\gamma\beta(1-\phi/\nu)x+\mu\right]x\label{eq:polyq}
\end{align}

\noindent are second order polynomials, i.e. parabolas. By assuming
instead $\dot{y}>0$ one obtains (\ref{eq:lisocline}) but with a
``>'' sign, which means that insect biomass grows for points lying
below the isocline and conversely decline for points above the isocline.

For function $p(x)$: $p(0)=-\eta\mu\nu<0$ and $\lim_{x\to\pm\infty}p(x)=+\infty$.
This means that $p(x)$ has one negative root and one positive root;
and also that $p(x)<0$ between the negative and positive roots, and
$p(x)>0$ otherwise. Since $p(x)$ is the denominator of (\ref{eq:lisocline}),
the larva isocline has the same roots as $p(x)$ in the x-axis. The
positive root of (\ref{eq:polyp}) and (\ref{eq:lisocline}) is:

\begin{equation}
x_{0}=\frac{\eta\nu}{2\epsilon\alpha}\left(1-\frac{\phi}{\nu}\right)+\sqrt{\left[\frac{\eta\nu}{2\epsilon\alpha}\left(1-\frac{\phi}{\nu}\right)\right]^{2}+\frac{\eta\mu\nu}{\epsilon\alpha\gamma\beta}}\label{eq:x0}
\end{equation}

For function $q(x)$: it has one root at $x=0$, a second one at:
\begin{equation}
x_{v}=-\frac{\mu}{\gamma\beta(1-\phi/\nu)}\label{eq:xv}
\end{equation}

\noindent and $\lim_{x\to\pm\infty}p(x)=-\infty$. This means that
$q(x)>0$ between $0$ and $x_{v}$, and $q(x)<0$ otherwise. Both
roots make the denominator of (\ref{eq:lisocline}) equal to zero,
which means that the larva isocline has two vertical asymptotes, $x=0$
and $x_{v}$.

And finally, the larva isocline has one horizontal asymptote:
\begin{equation}
y_{h}=\lim_{x\to\pm\infty}\frac{p(x)}{q(x)}=\frac{\epsilon\alpha}{\gamma\beta(1-\phi/\nu)}\label{eq:yh}
\end{equation}

Notice that the signs of $x_{v}$ and $y_{h}$ depend on $\phi/\nu$:

\begin{equation}
\begin{cases}
\phi<\nu: & x_{v}<0,y_{h}>0\\
\phi>\nu: & x_{v}>0,y_{h}<0
\end{cases}\label{eq:phi_vs_nu}
\end{equation}

This information about the parabolas $(p(x),q(x))$, and the signs
of the asymptotes $(x_{v},y_{h})$, is enough to sketch the possible
shapes of the larva isocline: the isocline crosses the x-axis at the
roots of $p(x)$; its jumps to infinity at the roots of $q(x)$; and
is positive (negative) whenever $p(x)$ and $q(x)$ have the same
(different) signs. According to (\ref{eq:phi_vs_nu}) we have two
main cases:
\begin{enumerate}
\item If $\phi<\nu$ the vertical asymptote $x_{v}$ is negative and the
horizontal asymptote $y_{h}$ is positive. As we can see, there are
two alternatives, depicted by Figure \ref{fig:app.liso}A and B. Both
are indistinguishable in the positive octant, which is the only part
that matters: they both start at the $x_{0}$ in the plant axis and
grow up to a plateau $y_{h}$.
\item If $\phi>\nu$ the vertical asymptote $x_{v}$ is positive and the
horizontal asymptote $y_{h}$ is negative. In this configuration we
also have two alternatives, as depicted in Figures \ref{fig:app.liso}C
or D. However, we can quickly dismiss alternative D: the insect is
meant to grow for points that are below the larva isocline, but since
the isocline is decreasing, this automatically means to grow when
plant abundance is low rather than high. This is nonsensical because
the plant always has a positive effect on insects.
\end{enumerate}
\begin{center}
\begin{figure}
\begin{centering}
\includegraphics[angle=-90,width=0.75\paperwidth]{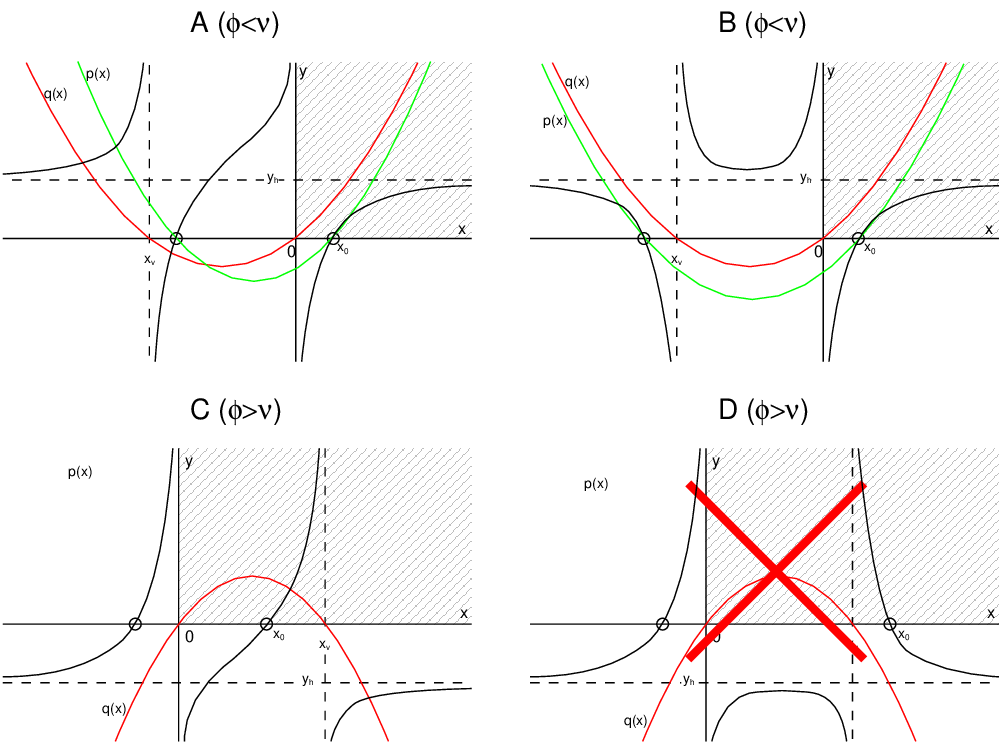}
\par\end{centering}

\protect\caption{\label{fig:app.liso}Possible configurations of the larva isocline,
pictured as a three segment black line. For A and B $\phi<\nu$. For
C and D $\phi>\nu$. Only parts in the positive octant (hatched square)
are considered. The green parabola $p(x)$ is the numerator of the
isocline and the circles indicate its roots, where $x_{0}$: positive
root. The red parabola $q(x)$ is the denominator of the isocline,
which has two roots $x=0$ and $x=x_{v}$, both of which are also
the vertical asymptotes of the isocline. The isocline also has an
horizontal asymptote $y_{h}$. The alternative in part D can be dismissed
because it implies a detrimental effect of plants on insects.}
\end{figure}

\par\end{center}

Figure \ref{fig:app.lisoshapes} shows how the positive part of the
larva isocline responds to some parameter changes. From the equations
that define the isocline's root (\ref{eq:x0}) and asymptotes (\ref{eq:xv},\ref{eq:yh})
we can conclude that increasing $\gamma,\beta$ tends to move the
isocline closer to the larva axis.

\begin{center}
\begin{figure}
\begin{centering}
\includegraphics[width=0.75\paperwidth]{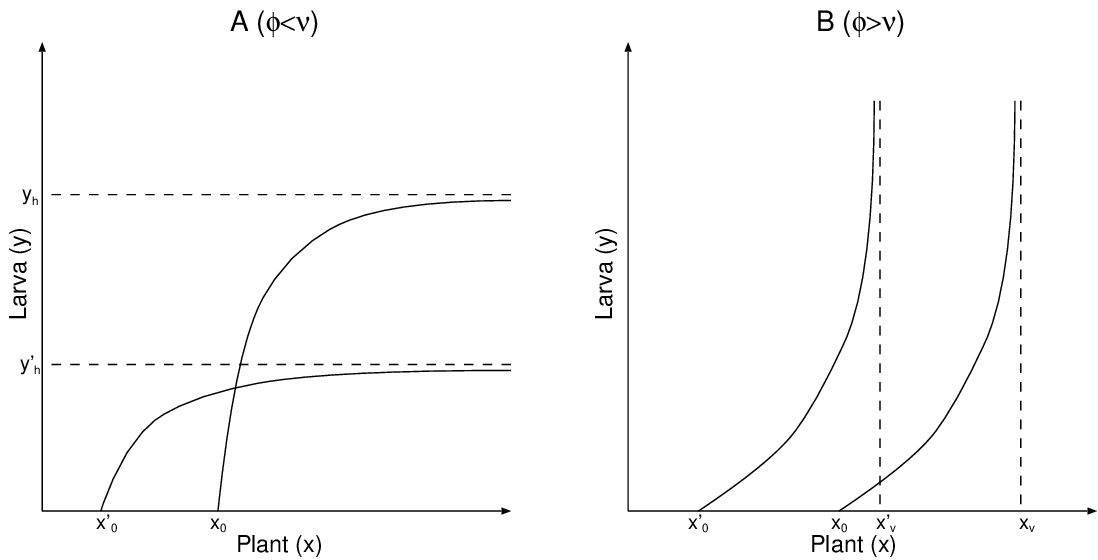}
\par\end{centering}

\protect\caption{\label{fig:app.lisoshapes} (A) For $\phi<\nu$ the larva isocline
moves closer to the larva axis and becomes more shallow as $\gamma$
and $\beta$ increase. For $\phi>\nu$ the larva isocline becomes
closer to the larva axis.}
\end{figure}

\par\end{center}

\subsection*{Appendix C: Additional simulations}

\addcontentsline{toc}{subsection}{Appendix C}
\renewcommand{\thefigure}{C.\arabic{figure}} \setcounter{figure}{0}

Figure \ref{fig:app.dynamics_cicle_above_k} displays limit cycles
in the PLA model with plant biomasses entirely above the carrying
capacity. The parameters are as in Table \ref{tab:vars_and_pars}
of the main text, but with $\gamma=0.00973,\beta=0.01$. Figure \ref{fig:app.dynamics_unscaled_ucycle}
shows an example where oscillations can damped out or evolve towards
a limit cycle depending on the initial conditions. Parameters as in
Table \ref{tab:vars_and_pars} of the main text, but with $\gamma=0.06,\beta=20,\nu=5$.
The attraction basins for both outcomes are separated by an unstable
orbit, like the one show in the bifurcation plot in Appendix A.

Figure \ref{fig:app.dynamics_pfla_unscaled} displays the dynamics
of plants, flowers, larva and adult insects under the interaction
mechanism (\ref{eq:pfla}) from which the PLA model is derived in
the main text. This simulation uses parameter values from the last
column of Table \ref{tab:vars_and_pars} with $\gamma=0.01,b=0.005$.
This figure is comparable to Figure \ref{fig:dynamics_muther} in
the main text: the 200 time in units there, become $t=\tau/r=200/0.05=4000$
time units here, and the plant's carrying capacity there $(x=1)$,
becomes $c^{-1}=0.01^{-1}=100$ here.

\begin{center}
\begin{figure}
\begin{centering}
\includegraphics[clip,width=0.75\paperwidth]{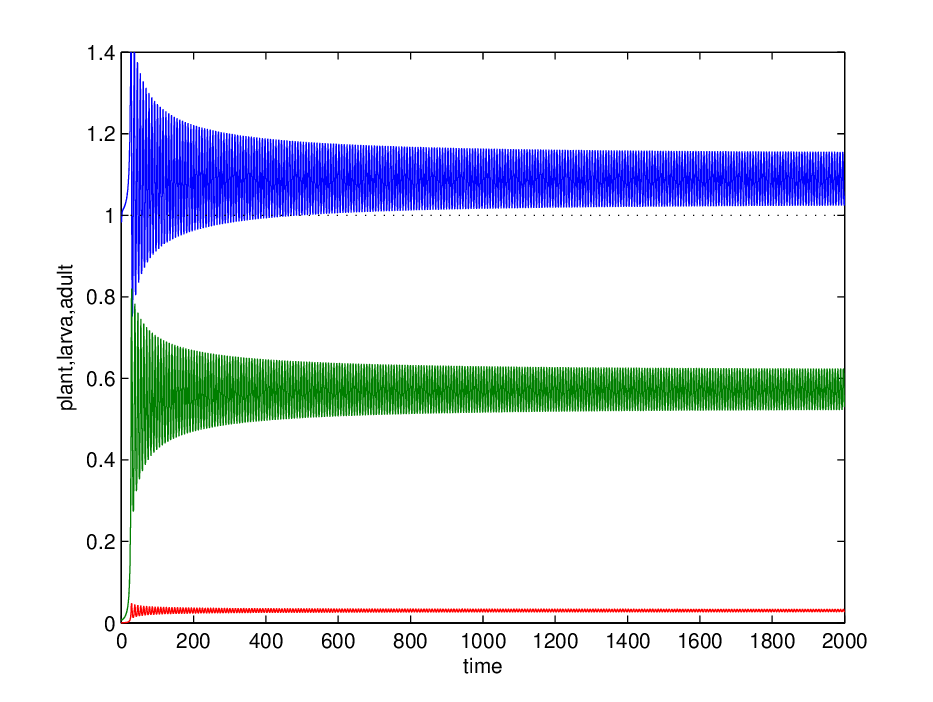}
\par\end{centering}

\protect\caption{\label{fig:app.dynamics_cicle_above_k}Limit cycles in the PLA model,
with plants above the carrying capacity (dotted line). Blue:plant,
green:larva, red:adult.}
\end{figure}

\par\end{center}

\begin{center}
\begin{figure}
\begin{centering}
\includegraphics[clip,width=0.75\paperwidth]{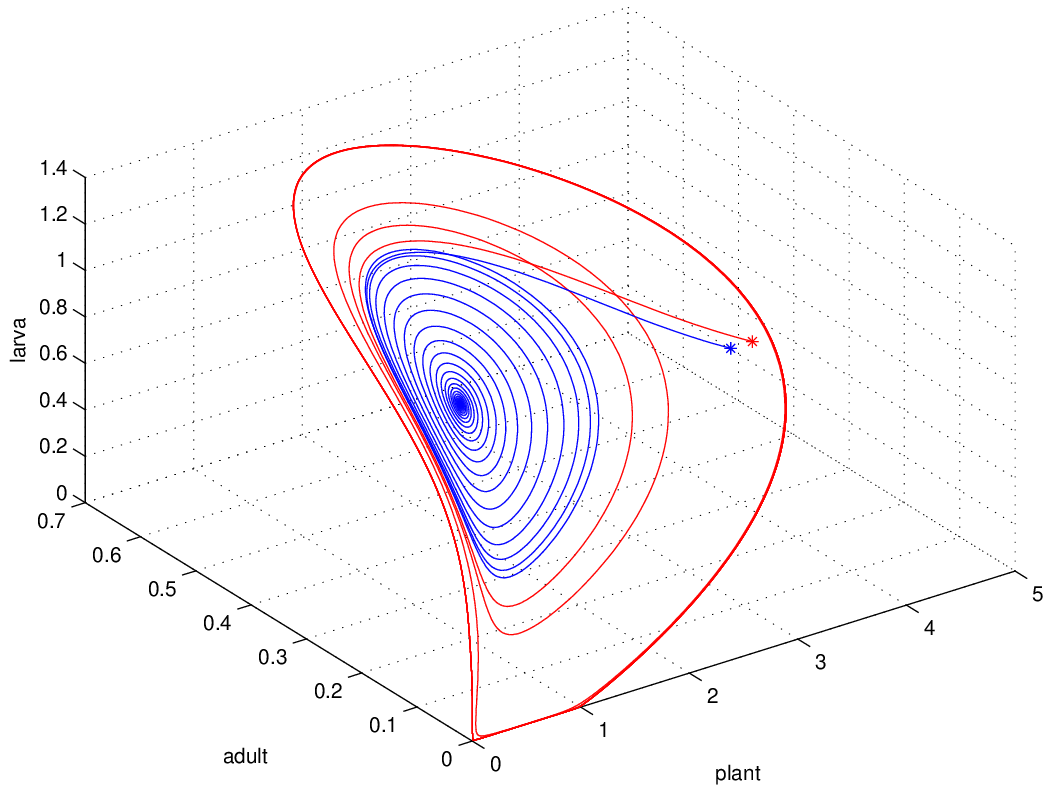}
\par\end{centering}

\protect\caption{\label{fig:app.dynamics_unscaled_ucycle}Oscillations in the PLA model
started with different initial conditions ({*}). The oscillations
can dampen out (blue) or converge to a limit cycle (red).}
\end{figure}

\par\end{center}

\begin{center}
\begin{figure}
\begin{centering}
\includegraphics[clip,width=0.75\paperwidth]{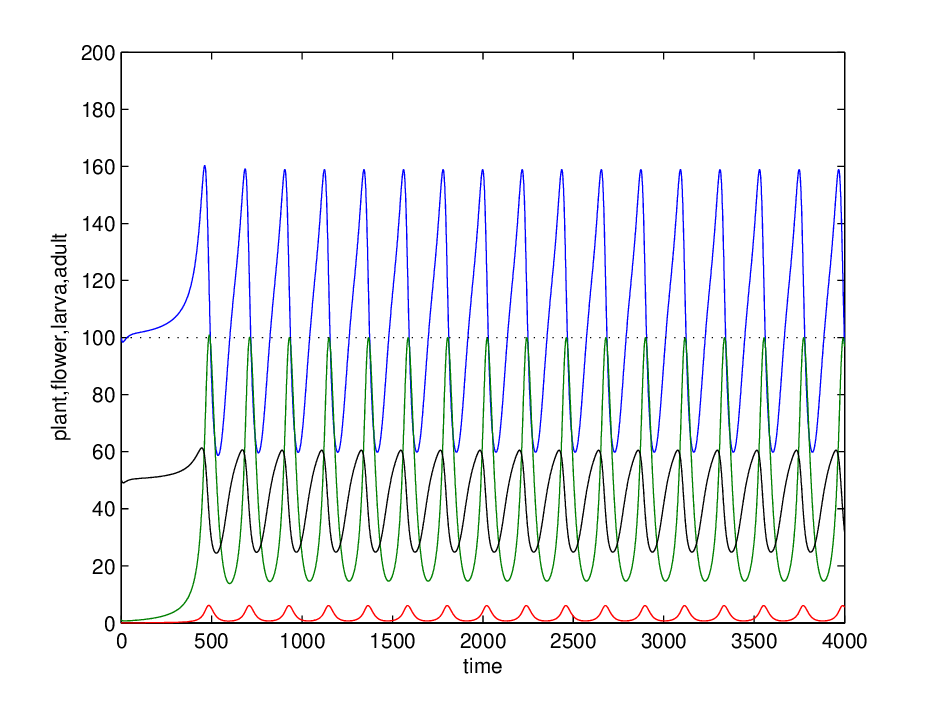}
\par\end{centering}

\protect\caption{\label{fig:app.dynamics_pfla_unscaled}Interaction dynamics of plants,
larva and adults, with the flowers explicitly considered. Blue:plant,
green:larva, red:adult, black:flowers. The dotted line indicates the
plant's carryng capacity.}
\end{figure}

\par\end{center}
\begin{lyxcode}
\end{lyxcode}

\end{document}